\input ./style/arxiv-general.cfg
\documentclass[aoas,MSNbibl,nameyear,dvips]{arximspdf}
\makeatletter
   \@ifpackageloaded{graphicx}{}{\usepackage{graphicx}}
\makeatother

\doi{10.1214/15-AOAS807}% Updated by VTEXPTS2LaTeX.exe, 12.02.2015
\volume{9}
\issue{1}
\pubyear{2015}
\firstpage{525}
\lastpage{546}
\docsubty{FLA}

\makeatletter
\def\cal{\mathcal}
\newcommand{\Norm}{\mathcal{N}}
\newcommand{\betavect}{\bolds{\beta}}
\newcommand{\phivect}{\bolds{\phi}}
\newcommand{\thetavect}{\bolds{\theta}}
\newcommand{\muvect}{\bolds{\mu}}
\newcommand{\xivect}{\bolds{\xi}}
\newcommand{\wvect}{\mathbf{w}}
\makeatother

\begin{document}
\begin{frontmatter}

\title{Bayesian nonparametric disclosure risk estimation via mixed effects log-linear models}
\runtitle{Nonparametric disclosure risk estimation}

\begin{aug}
% Corresponding author: Maurizio Filippone - maurizio.filippone@glasgow.ac.uk% Updated by VTEXPTS2LaTeX.exe, 13.02.2015 07:34
%maurizio.filippone@glasgow.ac.uk% Updated by VTEXPTS2LaTeX.exe,
%12.02.2015 14:55
\author[A]{\fnms{Cinzia} \snm{Carota}\thanksref{m1,T1}\ead[label=e1]{cinzia.carota@unito.it}},
\author[B]{\fnms{Maurizio} \snm{Filippone}\corref{}\thanksref{m2}\ead[label=e2]{maurizio.filippone@glasgow.ac.uk}},
\author[A]{\fnms{Roberto} \snm{Leombruni}\thanksref{m1,T1}\ead[label=e3]{roberto.leombruni@unito.it}}
\and
\author[C]{\fnms{Silvia} \snm{Polettini}\thanksref{m3,T2}\ead[label=e4]{silvia.polettini@uniroma1.it}}
\runauthor{Carota, Filippone, Leombruni and Polettini}
\affiliation{Universit\`a di Torino\thanksmark{m1}, University of
Glasgow\thanksmark{m2} and Universit\`a di Roma ``La
Sapienza''\thanksmark{m3}}

\thankstext{T1}{Supported in part from UniTo Project TO\_Call3\_2012\_0119 ``The Popart Network.''}
\thankstext{T2}{Supported in part from Sapienza University by Grant C26A14PNSC.}
% \and
\address[A]{C. Carota\\
R. Leombruni\\
Dipartimento di Economia e Statistica\\ % ``Cognetti de Martiis'' \\
Universit\`a di Torino \\
Lungo Dora Siena 100A\\
10153 Torino\\
Italy\\
% Lungo Dora Siena, 100A \\
% 10153 TORINO - Italy \\
\printead{e1}\\
\phantom{E-mail:\ }\printead*{e3}}
% \phantom{E-mail:\ }\printead*{e2}

\address[B]{M. Filippone\\
School of Computing Science\\
University of Glasgow\\
18 Lilybank Gardens\\
G12 8QQ, Glasgow\\
Scotland\\
% usually few lines long\\
\printead{e2}}

\address[C]{S. Polettini\\
Dipartimento di Metodi e modelli per \\
\quad l'economia, il territorio e la finanza \\
Sapienza Universita` di Roma\\
Via Del Castro Laurenziano 9\\
00161 Roma\\
Italy \\
\printead{e4}}

\end{aug}

% HISTORY:
%
\received{\smonth{6} \syear{2013}}% Updated by VTEXPTS2LaTeX.exe,
%12.02.2015 14:55
%
\revised{\smonth{10} \syear{2014}}% Updated by VTEXPTS2LaTeX.exe,
%12.02.2015 14:55

% ABSTRACT
%
\begin{abstract}
Statistical agencies and other institutions collect data under the
promise to protect the confidentiality of respondents.
When releasing microdata samples, the risk that records can be
identified must be assessed. To this aim, a~widely adopted approach is
to isolate categorical variables key to the identification and analyze
multi-way contingency tables of such variables. Common disclosure risk
measures focus on sample unique cells in these tables %are the number
%of sample uniques which are also population uniques and the expected
%number of correct guesses when each sample unique is matched with a
%subject randomly chosen from the corresponding population cell.
and adopt parametric log-linear models as the standard statistical
tools for the problem. Such models often have to deal with large and
extremely sparse tables that pose a number of challenges to risk estimation.
This paper proposes to overcome these problems by studying
nonparametric alternatives based on Dirichlet process random effects.
The main finding is that %, compared to traditional log-linear models
the inclusion of such random effects allows us to reduce considerably
the number of fixed effects required to achieve reliable risk estimates.
This is %extensively
studied on applications to real data, suggesting, in particular, that
our mixed models with main effects only produce roughly equivalent
estimates compared to the all two-way interactions models, and %so that
%our approach
are effective in
defusing potential shortcomings of traditional log-linear models.
This paper adopts a fully Bayesian approach that accounts for all
sources of uncertainty, including that about the population
frequencies, and supplies unconditional (posterior) variances and
credible intervals.
%Statistical agencies and other institutions that release data arising
%from sample surveys are obliged to protect the confidentiality of
%respondent's identities and sensitive attributes.
%In order to quantify the risk that individuals in the sample can be
%identified, a widely adopted strategy is to isolate categorical
%variables key to the identification and analyze multi-way contingency
%tables of such variables using parametric log-linear models.
%When employing parametric models, however, the only hope to capture
%the complexity of the problem is to consider contingency tables
%including key variables as well as all their two-way interactions.
%As a result, parametric models have to deal with extremely large and
%sparse design matrices that pose a number of computational challenges
%to their estimation.
%This paper proposes to overcome such problems by studying alternatives
%to parametric models based on Dirichlet process random effects.
%The main finding is that nonparametric models produce roughly
%equivalent, and sometimes more reliable, risk estimates compared to
%all two-way interactions parametric models.
%This is extensively studied on applications to real data, suggesting
%that the proposed nonparametric models are effective in reliably
%estimating risk and in defusing potential shortcomings of traditional
%log-linear parametric models.
%In this respect, the proposed nonparametric log-linear models are
%alternative to grade of membership models recently introduced by
%Manrique-Vallier and Reiter (2012).
\end{abstract}

% KEYWORDS
\begin{keyword}
\kwd{Bayesian nonparametric models}
\kwd{confidentiality}
\kwd{disclosure risk}
\kwd{Dirichlet process}
\kwd{log-linear models}
\kwd{mixed effects models}
\end{keyword}
\end{frontmatter}

%s1 #&#
\section{Introduction}\label{intro}
Statistical agencies and other institutions that release data arising
from sample surveys are obliged to protect the confidentiality of
respondent's identities and sensitive attributes.
%In releasing data arising from sample surveys statistical agencies and
%other organizations are obliged to protect the confidentiality of
%respondent's identities and sensitive attributes. %Confidential
%information may be disclosed if potential intruders can link records
%in the released data to other databases (that include direct
%identifiers) by matching on variables common to the two databases, or
%key variables. For most socio-demographic surveys the key variables
%are discrete and, assuming no errors in data sources, the problem of
%assessing disclosure risks associated with any proposed data release
%is tackled by: \\
In socio-demographic surveys the observed variables are often
categorical; some of these, called \textit{key variables}, are identifying
in that, %may also be
being also available in external databases, allow potential intruders
to disclose confidential information on records in the sample by
matching on such keys. Assuming that there are no errors in the
variables above, the problem of assessing disclosure risks associated
with any proposed data release is often tackled by:
(i) considering a contingency table representing the
cross-classification of subjects by the key variables (often this is a
very large and sparse table);
(ii) observing that a subject belonging to a cell with a sample
frequency of 1 (sample unique) is at a relatively high risk of
identification if there are few subjects %(records / statistical units)}
in the population with that
% corresponding population cell contains few subjects with that
%combination of values of the key variables.
combination
of the key variables. % {\it(if few subjects in the population share
%the same values of the key variables.)}

Common disclosure risk measures are the number of sample uniques which
are also population uniques and the expected number of correct guesses
when each sample unique is matched with a subject randomly chosen from
the corresponding population cell. Further measures can be found in
\citet{forster:webb} along with an extensive survey of the previous
literature; in this paper we selectively review only those references
%those aspects in the literature
that are
closely related to the focus of our work.

Disclosure risk is traditionally estimated by parametric models; in
this context, \citet{skinner:holmes}, \citet{fienb:makov:1998},
\citet
{carlson}, \citet{elamir:skinner}, \citet{forster:webb} and \citet
{skinner:shlomo} introduce a log-linear model for the expected cell
frequencies that overcomes the assumption of exchangeability of cells
of the contingency table, implying constant risk estimates across cells
having the same sample frequency. To learn about the risk in a given
cell from neighboring cells without relying on the association
structure implied by a log-linear model,
\citeauthor{rinott:shlomo} (\citeyear{rinott:shlomo,rinott:shlomo2})
%, instead,
propose a local smoothing polynomial model, %only
applicable to key variables for which a suitable definition of
closeness is available.
As far as estimation goes, the literature presents a whole variety of
strategies, including combinations of methods ranging from maximum
likelihood estimates to fully Bayesian estimates, and also a method
based on multiple imputation.

%Adopting a fully Bayesian approach, as in \citet{mvreiter:jasa12},
%also allows us to accout for uncertainty about population frequencies,
%that represents an additional source of variability of risk estimators.
Drawing from the above-mentioned literature, we propose a Bayesian
semi-parametric version of log-linear models, which specifically is a
mixed effects log-linear model with a Dirichlet process (DP) prior
[\citet{ferguson}] for modeling the random effects. As in \citet
{fienb:makov:1998}, \citet{forster:webb}, and
\citeauthor{mvreiter:jasa12} (\citeyear{mvreiter:jasa12,manriquevallier:reiter:13}), we adopt a fully Bayesian
approach. Unlike repeated sampling schemes, the Bayesian framework is
particularly appealing in a disclosure limitation context, where the
sample to be released is unique and fixed. It also allows us to account
for uncertainty about population frequencies,
%that
which thus represents an additional source of variability of risk
estimators. In this respect, our work is very different from previous
works based on log-linear models, including the one by \citet
{rinott:shlomovar}, as we provide unconditional variances and credible
intervals for sample disclosure risk measures.

Emphasizing the
%second
random effects component of the model, we will refer to it as a
\textit{nonparametric} log-linear model, its \textit{parametric} counterpart being
a log-linear model with random effects modeled parametrically; fixed
effects are always assigned a parametric prior, so no further
distinctions are necessary.\setcounter{footnote}{2}\footnote{The
reason for such and related abuses of terminology is that often in
the course of the paper we think of random effects conditionally on
fixed effects and vice versa. This is also why we refer to \emph{independence}
in the sequel.}
Our nonparametric log-linear models are special cases of the family of
hierarchical DPs [\citet{Teh06}] which also include some elements of the
class of mixed membership models [which in turn include grade of
membership models, \citet{erosheva2007}] such as latent Dirichlet
allocation models [\citet{Blei03}].
%In Section~3 the details of our estimation method are described. \\ In
%Section~\ref{appl} we compare parametric and nonparametric models
%based on a random sample extracted from the population defined by the
%Italian National Social Security Administration (2004), benchmarking
%risk estimates against the true values of risks. The same comparison
%is also provided by using a random sample from public use microdata
%from the state of California (IPUMS, Ruggles et al. 2010). \\
%The fully Bayesian estimation method adopted in this article %,
%described in Section~3, as in \citet[p. 1390]{mvreiter:jasa12}
%explicitly considers the randomness of
%Adopting a fully Bayesian approach, as \citet[p.
%1390]{mvreiter:jasa12}, allows us to accout for uncertainty about
%population frequencies representing an additional source of
%variability of risk estimators. In this respect, our work is very
%different from previous works based on log-linear models, including
%the one by \citet{rinott:shlomovar} on conditional variances and
%confidence intervals for disclosure risk measures. %We also remark
%that in a disclosure limitation context the sample to be released is
%unique and fixed and, unlike repeated sampling schemes, the Bayesian
%approach is particularly appealing as it works conditionally on the
%observed data.\\
%On the one hand,

The proposed %modeling assumption
nonparametric formulation
has two major advantages.
First,
% suitable specifications of the base measure of the DP allow for
%useful extensions of many parametric models; indeed, in Section~2 we
%generalize the ones adopted in \citet{skinner:holmes},
it may be interpreted as the nonparametric extension of some of the
parametric models proposed in the literature (see Section~\ref{sec:2}).
%in Section~4
%we apply to real data such model and a similarly genaralized version of
%sketch the analog of two further models proposed in
%one of the models proposed
Second, and most importantly,
%On the other hand,
%the assumption of DP random effects gives the modeling flexibility of
%accommodating any possible clustering of cells in the contingency
%table of the key variables, with cells in the same cluster receiving
%the same random effect. A practical consequence is that the huge
%number of patterns of dependence among cells automatically created by
%the proposed model may reduce the number of high-order terms required
%to achieve a satisfactory performance of risk estimators \citep[see,
%e.g.,][]{dorazio}. In particular,
in many applications to real data, two of which are presented in
Section~\ref{appl}, we observed roughly equivalent global %in some
%cases more reliable,
risk estimates under nonparametric log-linear models with main effects
only (say,
%``
nonparametric %
\textit{independence} models)
%''
compared to all two-way interactions log-linear models with and
without random effects.
Quoting \citeauthor{mvreiter:jasa12} [(\citeyear{mvreiter:jasa12}),
page 1390], the latter ``have been
found to produce reasonable results in many cases [\citet
{fienb:makov:1998,elamir:skinner,skinner:shlomo}], and so represent a
default modeling position.'' Consequently, our main finding is that
%The main finding in this article is that
our nonparametric independence models can be used as default models,
thereby avoiding %{\it(that, interestingly, are able to avoid)}
the severe difficulties associated with complex log-linear model
estimation in the presence of sparse tables
%without incurring the severe difficulties with complex models and
%sparse tables
%in log-linear model estimation
%due to (the) nonexistence of maximum likelihood estimators in the
%presence of complex models and sparse tables. These difficulties,
[see, e.g., \citet{fienberg2012}]. These difficulties arise
%in sparse tables
from certain patterns of %maximum likelihood estimators in very sparse
%tablew
sampling zeroes which
% are not fully informative about model parameters
make the model nonidentifiable
and result in nonexistent maximum likelihood estimators (MLE). This
fact has long been known [\citet{haberman:74}], but %it is now clear
recent research shows that nonexistent MLEs are likely to arise even in
small tables, in the presence of positive margins and in frequently
used models such as the all two-way interactions model. ``Under a
nonexistent MLE, the model is not identifiable, the asymptotic standard
errors are not well defined and the number of degrees of freedom
becomes meaningless''
[\citeauthor{fienberg2012} (\citeyear{fienberg2012}), page 997]. Moreover, common
statistical packages are inadequate to cope with this problem, as
detailed in %(a detailed analysis of standard procedures in SAS and R,
%a discussion of inferential consequences and further examples can be
%found in
\citet{fienberg2007}.
The issue of nonexistence of MLE is also important in Bayesian
analysis of contingency tables, but in our
%% (default ???)
nonparametric models it is defused in two ways. First, the only fixed
effects to be estimated are the main effects. This is a substantial
simplification of the log-linear model significantly reducing the
severity of the problem. Second, the vague prior we assign to fixed
effects replaces the information content lacking in the data with the
information contained in the prior about all cells. This obviates the
need for %This avoids the need to resort to
ad hoc additions of small positive quantities to cells containing
sampling zeroes [\citet{fienberg2007}, page 3437; \citet{fienberg2012},
page~1012], which is potentially severely misleading.

Recently, under the assumption that there are no structural zeroes in
the contingency table, \citet{mvreiter:jasa12} employ a Bayesian
version of the grade of membership model for disclosure risk
estimation, also discussing the model choice. This is a very
challenging problem in complex log-linear models only addressed in
\citet{skinner:shlomo}; another approach is Bayesian model averaging,
pursued by \citet{forster:webb} on decomposable graphical models.
In a subsequent paper, \citet{manriquevallier:reiter:13} propose a
truncated latent class model (LCM) for managing structural zeroes,
thereby removing a traditional limitation of Bayesian %versions of
latent structure models.
The paper is organized as follows: in Section~\ref{sec:2} we define our
model and interpret it in light of the existing literature; in
Section~\ref{inf} we describe in detail our estimation method. In
Section~\ref{appl} we compare parametric and nonparametric models based
on a sample extracted from the population defined by the Italian
National Social Security Administration
(WHIP-Work Histories Italian Panel,  Laboratorio Revelli, Centre
for Employment Studies, \url{http://www.laboratoriorevelli.it/whip}),
benchmarking risk
estimates against the true values of global risks. The same comparison
is also provided through a random sample from public use microdata from
the state of California [IPUMS, Ruggles et al. (\citeyear{ru10})].
In Section~\ref{sec:comput} we discuss comparisons between our nonparametric models
and the LCMs of \citet{manriquevallier:reiter:13}, showing %argue
that both rely on
%contain
the same basic assumptions, although implemented in different ways,
which leads to different models with relative merits over each other.
We also discuss some computational aspects, suggesting use of the
Empirical Bayesian version of our model to reduce the computational
burden for very large tables.
% [NON RIESCO A DIRLO IN MODO NON MACCHERONICO].
Finally, in Section~\ref{disc} we provide some final comments.
\section{%Semi-parametric
Log-linear models for disclosure risk estimation}
 %\label{log-lin}
\label{sec:2}
%In the contingency table of key variables we denote by

Let $f_k$ and $F_k$ denote the sample and population frequencies in the
$k$th cell, respectively, and let
%by
$K$ be the total number of cells in the contingency table of the key
variables. Our goal is to estimate global risks of re-identification,
or disclosure risks, defined as
%
%e1 #&#
\begin{equation}
\label{eq:tau1} \tau_1=\sum_{k=1}
^K I(f_k=1, F_k=1)= \sum
_{k=1} ^K I(f_k=1) \tau_{1,k},
\end{equation}
that is, the number of sample uniques which are also population
uniques, and
%
%e2 #&#
\begin{equation}
\label{eq:tau2} \tau_2=\sum_{k=1}^K
I(f_k=1)\frac{1}{F_k}=\sum_{k=1}
^K I(f_k=1) \tau_{2,k},
\end{equation}
that is, the expected number of correct guesses if each sample unique
is matched with an individual randomly chosen from the corresponding
population cell [see, e.g., \citet{rinott:shlomo}].
Usually
%, under the assumption of cell independence,
these measures are approximated by their expectations $E(\tau
_i|f_1,\ldots,f_K)$, $i=1,2$,
%under the assumption of cell independence,
namely, under the assumption of cell independence,
%
%e3 #&#
\begin{eqnarray}
\label{tau1*} \tau_1^*&=&\sum_{k=1}
^K I(f_k=1) \operatorname{Pr}\{F_k=1|f_k=1\}=
\sum_{k=1} ^K I(f_k=1)
\tau_{1,k}^*,
\\
\label{tau2*} \tau_2^*&=&\sum_{k=1}^K
I(f_k=1)E(1/F_k|f_k=1)=\sum
_{k=1}^K I(f_k=1) \tau_{2,k}^*,
\end{eqnarray}
and estimated by using parametric models, which are often elaborations
of the Poisson model. Assuming
% \begin{equation}
% \label{poisson}
% F_k\sim Poisson(\lambda_k) f_k\sim Poisson(\pi\lambda_k) \mbox{
%independently } k=1,\ldots,K,
% \end{equation}
%e5 #&#
\begin{equation}
\label{poisson} F_k\sim\mathrm{Poisson}(\lambda_k)
\quad\mathrm{and}\quad f_k\sim \mathrm{Poisson}(\pi\lambda_k)
\end{equation}
independently for $k=1,\ldots,K$, with
$\pi$ being the (known) sampling fraction, the terms in (\ref{tau1*})
and (\ref{tau2*}) can be expressed in closed form,
%
%e6 #&#
\begin{equation}
\label{tau*:cella} {\tau}_{1,k}^*=e^{-(1-\pi)\lambda_k},\qquad {\tau}_{2,k}^* =
\bigl(1 - e^{-(1-\pi)\lambda_k} \bigr)/ \bigl((1-\pi)\lambda_k \bigr).
\end{equation}
%
% \maurizio{tolgo Among these}
%The work of Bethlehem \textit{et al.} %(1990)
%statistical model for samples where the identifying variables form a
%contingency table.
%the model is a hierarchical Poisson-Gamma superpopulation model where
%$F_k \sim Poisson (\lambda_k)$ and $f_k | F_k \sim{binomial} (F_k ,

%F_k | \gamma_k & \sim& \mbox{Poisson} (N \gamma_k), \\
%f_k | F_k , \gamma_k , \pi& \sim& \mbox{binomial} (F_k , \pi),
%in which $\gamma_k$ is the probability that a unit of the population
%falls into cell $k$ and $\pi$ is the sampling probability.
%The model was used to deduce (\ref{eq:tau1}) and can be seen as an
%approximation to the Dirichlet-multinomial model analysed by Takemura
%%(1998)
%
%estimation}
% Always give a unique label
% and use \ref{<label>} for cross-references
% and \citet{<label>} for bibliographic references
% use \sectionmark{}
% to alter or adjust the section heading in the running head
%models in the disclosure literature are parametric}.
%Although
In a relevant part of the literature the Poisson assumption is
integrated by log-linear modeling of cell means %to avoid unrealistic
%exchangeability assumptions. Different parametric log-linear models
%have been explored
and, as mentioned in Section~\ref{intro}, the all two-way interactions
model %, that does not include
without random effects has been recognized as a useful default model
by many authors [\citet{fienb:makov:1998,elamir:skinner,skinner:shlomo}]; recent articles, however, show that inference in
this model is not trivial with sparse tables. Even if the parameters of
interest are the cell means $\lambda_k$, and the Iterative Proportional
Fitting (IPF) is guaranteed to converge to the extended MLE %[NOTO CHE
%RINALDO AMA INDUGIARE NEL BY DESIGN. potremmo scriver by construction
%che ha pi{\chr"F9} senso anche se la frase {\chr"E8} presa da three
%centuries] by design
by construction, the rate of convergence, with the noticeable
exception of decomposable graphical models,
%(which is linear when the MLE exist)
can be very slow when the MLE is not defined. In conclusion, ``The
behavior of IPF when the MLE does not exist has not been carefully
studied to date'' [\citet{fienberg2007}, page 3438]. %All these facts
The previous facts, along with the nature of the problem, motivate our
attempt to address it in a Bayesian nonparametric framework by
introducing DP random effects. The assumption of a DP prior gives the
modeling flexibility of accommodating any possible clustering of cells
in the contingency table of the key variables, and implies that all
possible clusters of cells are considered,
with cells in the same cluster receiving the same random effect.
A practical consequence is that the huge number of patterns of
dependence among cells automatically created by the DP prior may reduce
the number of high-order terms required in the log-linear model to
achieve a satisfactory performance of risk estimators [see, e.g., \citet
{dorazio}]. Aiming at exploring this idea in real data applications, we
build on work by \citet{skinner:holmes} and related papers, such as
\citet{elamir:skinner} and \citet{carlson}. Before describing our
proposal, we briefly review the above references. %, and propose the
%nonparametric extensions of their models that will be described later
%in this Section. We start with a brief review of the approach by
%Their estimation strategy will be extended in section 3.
%their performances in disclosure risk estimation to the performances
%of the undelying parametric models.
% and related work, such as \citet{elamir:skinner} and \citet{carlson}
%As mentioned, many relevant models in the disclosure literature are
%parametric, and rely on Poisson
%log-linear modelling of cell means to avoid unrealistic
%exchangeability assumptions; the difficulties that arise in log-linear
%analysis of sparse tables with sampling zeroes that are not fully
%informative about the model parameters motivate our attempt to addres
%the issue in a Bayesian nonparametric context.
%, specifically extending the model and the estimation strategy
%introduced by \citet{skinner:holmes}.We build on work by
%defined in their paper and related work, such as
%estimation procedure by proposing a fully Bayesian approach to
%estimating (\ref{eq:tau1})-(\ref{tau2*}) that accounts for all sources
%of uncertainty.
%to models s work by \citet{skinner:holmes} to derive a nonparametric
%extension of models defined in their paper and related work, such as
%Since in Section~3 we also extend their estimation procedure (by
%proposing a fully Bayesian approach to estimating (\ref{eq:tau1})-(
% First, we briefly review their work. \\
%First, we briefly review their work. %in \citet{skinner:holmes}
%and then report the proposed nonparametric extensions. %Both are
%quickly sketched in the next two paragraphs. \\
Assuming (\ref{poisson}), %$F_k\sim Poisson(\lambda_k)$ and $f_k\sim
%Poisson(\pi\lambda_k)$ independently for $k=1,\ldots,K$, %Skinner and
%Holmes (1998)
Skinner and Holmes %the work in \citet{skinner:holmes}
model the parameters $\lambda_k$ through a log-linear model with mixed
effects:
%
%e7 #&#
\begin{equation}
\lambda_k=e^{\mu_k},\qquad \mu_k= {
\wvect'_k} \betavect+ \phi_k, \label{formula2}
\end{equation}
where $\wvect_k$ is a $q\times1$ design vector depending on the values
of the key variables in cell $k$, $  \betavect$ is a $q\times1$
parameter vector (typically main effects and low-order interactions of
the key variables), and $\phi_k$ is a random effect accounting for
cell-specific deviations. % The sampling fraction $\pi$ is supposed to
%be known.
% Finally, as regards the distribution of the random effects,
Finally,
%as far as the distribution of random effects goes, Skinner and Holmes
$\phi_k\   \mathrm{i.i.d.} \sim   \Norm(0,\sigma^2)$.
Formula~(\ref{formula2}) can be re-expressed using multiplicative
random effects as $ \lambda_k=e^{{\wvect'_k}\betavect} e^{\phi
_k}=\xi_k
  \omega_k$, hence %therefore we obtain
$\lambda_k|(\betavect, \sigma^2)\sim   \mathrm{Lognormal}(\wvect'_k
\betavect,\sigma^2)$, independently for $k=1,\ldots,K$.

\citet{skinner:holmes} estimate $\tau_1^*$ of formula~(\ref{tau1*}) by
%%In all three cases, this is
a two-stage procedure: in the first stage, the association among cells
is exploited to estimate the hyperparameters $\betavect$ and $\sigma^2$
of the Lognormal prior; % while,
in the second (and completely separate) stage, estimates of ${\tau
}_{1,k}^*$ are obtained cell by cell, independently.
When the preliminary estimate of $\sigma^2$ is positive, this procedure
leads to empirical Bayes estimates of the ${\tau}_{1,k}^*$'s in (\ref
{tau*:cella}), otherwise the random effects $\phi_k$'s are removed, and
plug-in estimates of the $\tau_{1,k}^*$'s are derived by using ML
estimates $\hat\xi_k=e^{\wvect'_k\hat{\betavect}}$.
In the same framework, but focusing on estimation of $\tau_{2,k}^*$ in
(\ref{tau*:cella}), \citet{elamir:skinner} assume independent Gamma
priors in place of Lognormals on $\lambda_k$'s, and find that the
addition of parametric random effects does not improve risk estimates;
%%are unnecessary to improve risk estimates;
%obtained by all two way interactions models are not improved by
%considering parametric random effects;
%are unecessary to
%do not improve the global risk estimates obtained by all two way
%interactions models;
as a consequence, they %also
suggest to adopt %they thus adopt
%thus adopting
%thus they adopt
%similar
plug-in estimates. %of the $\tau_{2,k}^*$'s in (\ref{tau*:cella}).
%%Based on this finding,
%More recently,
%searching for a log-linear model without random effects that is
%optimal in terms of risk estimation. Consequently, in the above
%mentioned approaches,
%so that
%${\tau}_{1,k}^*$ and ${\tau}_{2,k}^*$ are estimated by plug-in.
%is always estimated by equation~(\ref{eq:plugin}), and similarly a
%plug-in estimate is used for the summands in $\tau_2^*$, i.e. ${
% Actually, the inclusion of parametric random effects in the
%log-linear model is shown to be ineffective (fruitless) in
%priors of $\lambda_k$'s are taken to be independent Gamma
%distributions instead of Lognormal.
Conjugate Gamma priors guarantee computational advantages, as do the
Inverse Gaussian distributions ($\operatorname{IG}$) described in \citet{carlson}.

Our proposal is as follows: %building on work by \citet{skinner:holmes},
we keep the mixed effects log-linear structure (\ref{formula2}), but
remove the assumption of normality.
% fiwe
%remove the assumption of normality and
We model the distribution function $G$ of the random effects as unknown
and a priori distributed according to a DP $\cal{D}$ with base
probability measure $G_0$ and total mass parameter $m$ [\citet{ferguson}],
%
%e8 #&#
\begin{equation}
\phi_k|G \sim \mathrm{i.i.d.}\ G,\qquad G \sim{\cal{D}}(m, G_0).
\end{equation}
%
%where $G_0=N(\alpha, \sigma^2)$. % Further generalization are possible
%by selecting different specifications for $G$, for instance setting
%$G_0=Gamma(a,b)$ amounts to extending the model defined in Elamir and
%Skinner. \cinzia{avevo pensato anch'io di scriverlo qui, ma poi
%ripiegato per la citazione che vedi sopra perche' e' Gamma la
%distribuzione dei $\lambda$ e non dei $\phi$.}
Since $E(G)=G_0$ and $m$ controls the variance of the process, in
practice, $G_0$ specifies one's ``best guess'' about an underlying
model of the variation in $\phi$, and $m$ specifies the extent to which
$G_0$ holds.
Within the class of models just defined, we consider three
specifications of $G_0$ that lead to three different direct
generalizations of the existing literature, namely, \citet
{skinner:holmes}, when $G_0=N(\alpha, \sigma^2)$; \citet{carlson}, when
$G_0=\operatorname{IG}(\alpha, \sigma^2)$; \citet{elamir:skinner}, when $G_0=\operatorname{LG}(a,b)$,
%%$G_0=\operatorname{LG}(\alpha, \sigma^2)$,
where $\operatorname{LG}$ denotes the distribution of a log transformation of a
$\operatorname{Gamma}(a,b)$ variate $\omega$, with $f(\omega; a, b) =b^a / \Gamma(a)
\omega^{a-1} e^{-b \omega}$. %assumed
%a log transformation of a
% Gamma distributed with shape $a$ and rate $b$.
%whose mean and variance are $\alpha$ and $\sigma^2$, respectively
%(PENSO SIA MEGLIO PARAMETRIZZARE LA Log-Gamma COME UNA Gamma, CIOE'
%UTILIZZARE a=shape, b=rate).
% \citet{carlson}, when $G_0=\operatorname{IG}(\alpha, \sigma^2)$;
%Two distinct generalizations of the Skinner and Holmes model are
%presented, based on different specifications of the model parameters.
%In the first extension, our prior on $\betavect$ degenerates at $
%vector.
%This extension is directly inspired by both the structure of the
%model and the estimation strategy in \citet{skinner:holmes}.
%Therefore, the corresponding risk estimates will be referred to as {
%a generalization of (\ref{eq:EB}).
The hyperparameters %$(\alpha, \sigma^2)$
in the base measure $G_0$ can be fixed, which is how we proceed, or be
given a prior distribution. While in the corresponding parametric
approaches %of \citet{skinner:holmes}, \citet{elamir:skinner} and
a fixed distribution $G=G_0$ is selected and its hyperparameters are
estimated, we take the opposite perspective, that is, we assume a
random $G$ while holding the hyperparameters of its mean distribution
$G_0$ fixed, and chosen so as to obtain a vague specification. % $G_0$.
%; in the case when $\alpha=0$ we simply have some variability around
%the normal model assumed by Skinner and Holmes.
% In the second extension, we add the uncertainty about $\betavect$.
%To overcome identifiability issues, following \citet{li:mueller:lin},
%we drop the overall effect $\beta_0$, referred to as the ``intercept
%term'', from $\betavect$ % the fixed effects in (\ref{formula2}),
% and attempt to infer $\alpha$ in $G_0$ instead. %, and denote the
%rest of the $\beta$'s by $b$.
%A similar extension of the models in \citet{elamir:skinner} or
%probability base measure $G_0$.
%****QUI PARTE DI CINIZIA SULLA FLESSIBILITA' del DP

The estimation of risk measures under the proposed model is discussed
in Section~\ref{inf}. Here we analyze the implications of our
nonparametric specification of random effects and the advantages over
the parametric counterparts of our model. The clustering induced by the
DP prior on the random effects can be seen from a Polya-urn scheme
representation of the joint distribution of realizations from ${\cal
{D}}(m, G_0)$.
\citet{Blackwell73} provide this as the product of successive
conditional distributions:
% having the following form:
%e9 #&#
\begin{equation}\quad
\label{eq:polyaurn} \phi_i|\phi_1, \ldots,
\phi_{i-1},\qquad M \sim\frac{m}{m+i-1} G_0(\phi
_i)+\frac{1}{m+i-1}\sum_{k=1}^{i-1}
\delta(\phi_k=\phi_i),
\end{equation}
with $\delta(\cdot)$ denoting the Dirac delta function.
The above representation shows that clusters in the $K$ cells of the
population contingency table are induced by the existence of a positive
probability that a newly generated $\phi_i$ coincides with a previous
one. %Moreover, it
It also shows that $m$, the mass or precision parameter of the DP,
affects the expected number of clusters.

Therefore, under the previous assumptions,
%in the more general case
the likelihood function turns out to be a sum of terms where all
possible partitions (clusterings) $C$ of the $K$ cells into $c$
nonempty clusters are considered [see, e.g., \citet{lo,liu}],
%
%e10 #&#
\begin{equation}
\label{likelihood}% L(b|\mathbf{f})=
\sum_{c=1}^K
\sum_{C:|C|=c} \frac{\Gamma(m)}{\Gamma(m+K)} m^c \prod
_{j=1}^c \Gamma(n_j) \int p(
\mathbf{f}_{(j)}| \betavect, \phi_j)\,dG_0(
\phi_j),
\end{equation}
where $\mathbf{f}=f_1,\ldots,f_K$ and $n_j$ ($1\leq n_j \leq K$)
denotes the number of cells in the $j$th cluster,
%
%e11 #&#
\begin{equation}
\frac{\Gamma(m)}{\Gamma(m+K)} m^c \prod_{j=1}^c
\Gamma (n_j)=\operatorname{Pr}\{n_1,\ldots,n_c|C, c \},
\end{equation}
and finally
%
%e12 #&#
\begin{equation}
\label{lik:cluster} p(\mathbf{f}_{(j)}|\betavect, \phi_j)=\prod
_{k\in\mathrm
{cluster}
j  } \frac{1}{f_k!} e^{\pi f_k(\wvect'_k \betavect+ \phi_j)}
e^{-e^{\pi(\wvect'_k \betavect+\phi_j)}}.
\end{equation}
In the likelihood, starting from the latter formula, we %can observe
notice that the same random effect is assigned
%within each cluster of different partitions of the $K$ cells in an
%unknown number of clusters.\\
to all cells belonging to the same cluster, that is, to $\mathbf
{f}_{(j)}$, that $\operatorname{Pr}\{n_1,\ldots,n_c|C, c \}$ is the multivariate Ewens
distribution (MED) of $K$ distinguishable objects, or cells $\{1,\ldots
,K\}$ [see \citet{takemura};
\citet{Johnson}, Chapter~41], and that the number
of clusters in each %such partitions
partition ranges from 1 to $K$. We stress that the total number of
terms in the likelihood~(\ref{likelihood}) is the Bell number, $B_K$,
which is a combinatorial quantity assuming large values even for
moderate $K$; just to fix ideas, when $K=10$, $B_K=115\mbox{,}975$. %The model
%by Skinner and Holmes (1998)
The parametric counterparts of our nonparametric random effects models
correspond to just one term (namely, $c=K$) in the likelihood and,
consequently, even for moderate values of $K$, our model implies a huge
number of additional patterns of dependence among cells.

%MODIFICA SUGGERITA DA SILVIA
The above considerations show that the intrinsic
%distinctive
characteristics of DP random effects set them apart from parametric
random effects for their potential to improve upon the fixed effects
component of the log-linear %component of the
model.
Indeed, the fixed effects included in the log-linear model imply
specific patterns of dependence among cells. For instance, an
independence model implies that inference on a given cell depends on
all cells sharing a value of a key variable with it, since the
sufficient statistics are given by the marginal counts. The addition of
independent parametric random effects, $N(\alpha, \sigma^2)$,
$\operatorname{IG}(\alpha
, \sigma^2)$ or $\operatorname{LG}(a, b)$, %$\operatorname{LG}(\alpha, \sigma^2)$,
allows for departures from the Poisson log-linear model such as
overdispersion, but does not significantly affect the way one can learn
about a given cell from other cells. %On the contrary,
In contrast, the inclusion of DP random effects implies that, in
addition to the above-mentioned fixed effects
%log-linear
patterns, the model encompasses all other nonempty subsets of the $K$
cells. %, and that,
For each given partition, a possible relation of dependence among cells
in the same subset %(i.e. whether or not one can learn from those cells
%about any fixed cell in such a subset)
is explicitly evaluated. In other words, to learn about a given cell,
additional information is borrowed from cells belonging to the same
subset, for each subset to which the cell can be assigned in the
context of all possible partitions in nonempty subsets of the $K$ cells.
%;
This suggests both the potential for the proposed model to improve the
risk estimates and the associated computational complexity.
Furthermore, the results under our nonparametric models can be
interpreted as averages over mixed effects log-linear models with
different clusterizations of parametric random effects.
\section{Inference}\label{inf}

In this section we describe how to estimate not only $\tau_1^*$ and
$\tau_2^*$ in (\ref{tau1*}) and (\ref{tau2*}), as most of the
literature based on log-linear models does, but also $\tau_1$ and
$\tau
_2$ and their terms $\tau_{1k}$ and $\tau_{2k}$ in (\ref{eq:tau1}) and
(\ref{eq:tau2}), in a fully Bayesian way. This approach is inspired by
\citeauthor{mvreiter:jasa12} (\citeyear{mvreiter:jasa12,manriquevallier:reiter:13});
see also
\citet{fienb:makov:1998}. In order to keep the notation uncluttered,
let $\thetavect$ denote the set of
all
model
parameters conditioning $\lambda_1,\ldots,\lambda_K$ for each of the
models analyzed in this article.
%%except for $F_1,..,F_K.$ Denoting by $\thetavect$ the set of
The posterior distribution over $\thetavect$ is not available in closed
form for any of the models considered here. We employ Markov Chain
Monte Carlo (MCMC) techniques [\citet{Neal93}] to obtain samples from
$p(\thetavect| f_1, \ldots, f_K)$;
%Rather than focusing on $\tau_1^*$ and $\tau_2^*$, in this paper we
%directly estimate $\tau_1$ and $\tau_2$ and their summands $\tau_{1k},
%Note that $\tau_{ik}=\tau_{ik}(f_k,\lambda_k,F_k)   (i=1,2)$ %are
%functions of $f_k$, $\lambda_k$ and $F_k $
% where %$\tau_i=\tau_i(F_1,\ldots,F_K, \mathbf{f})$, $i=1,2$,
%$F_1,\ldots,F_K$ are unobservable random quantities (parameters), with
%$F_k|\lambda_k \sim Poisson(\lambda_k)$, $k=1,\ldots,K$. To perform
%posterior inference, we consider values of $\lambda_k$'s drawn from
%their joint posterior distribution and then values of $F_1,\ldots,F_K$
%drawn from the corresponding Poisson distributions. The relation $
%a sample of $\tau_{ik}$, $i=1,2$, from which it is possible to
%characterize the posterior distribution of global and cell-specific
%risk values by standard Monte Carlo techniques.
in particular,
%we denote by $\thetavect$ the vector of all parameters involved in
%such a posterior distribution, $\thetavect= \betavect, \phivect, m,
we propose to use a Gibbs sampler where we sample one group of
parameters at a time, namely, $\betavect| \mathrm{rest}$, $\phivect|
\mathrm{rest}$ and $m | \mathrm{rest}$.
The proposed Gibbs sampler steps are {briefly} discussed next.

\textit{Sampling} $\betavect$.
Given the form of the Poisson likelihood, it is not possible to sample
$\betavect$ using an exact Gibbs step, and so-called Metropolis within
Gibbs samplers need to be employed, whereby a proposal is accepted or
rejected according to a Metropolis ratio
[\citet{Roberts09}]. % samplers such as Metropolis-Hastings
%name a few, and aim at updating a set of variables at a time.
Recent work shows that it is possible to efficiently sample from the
posterior distribution of parameters of linear models using so-called
\textit{manifold MCMC} methods. %
Briefly, such samplers exploit the curvature of the log-likelihood
$\log
[p(f_1, \ldots, f_K | \betavect, \mathrm{rest})]$ by constructing a
proposal mechanism on the basis of the Fisher Information matrix [see
\citet{Girolami11} for further details].
In this work we adopt a Simplified Manifold Metropolis Adjusted
Langevin Algorithm (SMMALA) to sample $\betavect$ as previously done in
\citet{FilipponeDiscussionSMA11}, which simulates a diffusion on the
statistical manifold characterizing $p(f_1, \ldots, f_K | \betavect,
\mathrm{rest})$.
Define $M$ to be the metric tensor obtained as the Fisher Information
of the model plus the negative Hessian of the prior, and $\varepsilon$ to
be a discretization parameter.
SMMALA is essentially a Metropolis--Hastings sampler, with a
position-dependent proposal {akin to the Newton method in
optimization}, $p(\betavect^{\prime} | \betavect) = N(\betavect
^{\prime
} | \muvect, \varepsilon^2 M^{-1})$, with $\muvect= \betavect+ \frac
{\varepsilon^2}{2} M^{-1} \nabla_{\betavect} \log[p(f_1, \ldots, f_K |
\betavect, \mathrm{rest})]$.
Gradient and metric tensor can be computed in linear time in the number
of cells $K$ and in cubic time in the size of $\betavect$; therefore,
the method scales well to large data sets, but it may be
computationally intensive for highly parameterized models.

\textit{Sampling} $\phivect$.
An extensive treatment of MCMC for DP models can be found in \citet
{Neal00}, where we refer the reader for full details.
Drawing samples from the posterior distribution over the random effects
entails allocating cells to an unknown number of clusters and drawing a
value for the random effect for each cluster.
The way in which these steps are carried out depends on whether it is
possible to exploit conjugacy of the base measure, that is, whether the
integral $ \int p(f_k|\bolds\betavect, \phi) \,dG_0(\phi)$ can be
evaluated analytically.\footnote{Note that here $p(f_k|\bolds
\betavect
, \phi)$ represents the likelihood based on a single datum, that is, one
of the terms in the product (\ref{lik:cluster}).}

%QUI SI POTREBBE TOGLIERE: E' UNA RIPETIZIONE
%
%If
%When
%this is the case, %in the procedure to allocate
%when allocating cells to clusters the realizations of the random
%effects can be analytically integrated out, and realizations of the
%random effects can be obtained directly by a Gibbs step, as %indicated
%below.
%Otherwise, the procedure to allocate cells to clusters remains
%dependent on the particular realizations of the random effects, and a
%Metropolis-Hastings step must be introduced to update %them, leading
%to algorithms with slow convergence.
% % % % % %TOLTO FIN QUI
% In all cases, we consider Gibbs sampling schemes where we draw from
%the full conditional distribution of the random effect for each cell
%in the table, given the rest, and make use of the clustering structure
%induced by the DP.
In the applications presented in Section~\ref{appl}, we %extend the
%model by \citet{elamir:skinner} %and
choose a $\operatorname{LG}$ distribution for $G_0$ so that $\omega=e^{\phi}$ is given
a Gamma base measure. In this case we can exploit conjugacy with the
Poisson likelihood; a similar argument applies when $\phi$ is given the
$\operatorname{IG}$ distribution, for which the integral above is analytically tractable.
When conjugacy holds, a simple and efficient algorithm can be
constructed to draw samples from the full conditional distribution over
the random effects, which is referred to as Algorithm~3 in \citet{Neal00}.
First, the allocation of cells to clusters is updated for one cell at a
time, integrating out analytically the dependency from the actual value
that the random effects can take, and allowing the total number of
clusters to vary across iterations.
Second, the value of the random effect pertaining to each cluster can
be drawn directly from a known distribution [which is a Gamma in the
extension of \citet{elamir:skinner}], again due to the fact that the
% single datum
likelihood and the DP base measure form a conjugate pair.
The sampling of $\phivect$ has a computational cost that scales
linearly with the number of cells.

Instead, when we extend the model proposed by \citet{skinner:holmes},
the normal distribution does not enjoy the above-mentioned conjugacy
property; for this reason, %Algorithm~5 or other
sampling schemes for nonconjugate base measures described, for example,
in \citet{Neal00}, must be employed, and these usually lead to less
efficient MCMC sampling schemes.
\textit{Sampling} $m$.
In the literature, it has often been reported that inference in models
involving DPs is heavily affected by the mass parameter $m$, and that
setting it by means of Maximum Likelihood is bound to yield poor
results [see, e.g., \citet{liu}].
Rather than fixing this parameter, we propose to draw samples from its
posterior distribution and to account for uncertainty about it when
inferring risk measures.
% In order to do that, we log-transform $m$ and sample $\psi_m =
%avoid the Metropolis step, the approach of \citet{Escobar:West} could
%also be employed.
By selecting a Gamma prior over $m$, it is possible to employ the
approach of \citet{Escobar:West} to draw samples from the posterior
distribution over $m | \mathrm{rest}$ directly.
% ******* QUESTI LI HO FISSATI
%Sampling the hyperparameters of the base measure of the Dirichlet
%process proceeds as in standard Bayesian hierarchical models. When we
%choose a Gaussian base measure for the random effects, by imposing a
%Gaussian prior on the mean $\alpha$ and an Inverse Gamma prior on the
%variance $\sigma^2$ of the base measure, we can exploit conjugacy and
%obtain the conditional distribution of $\alpha$ and $\sigma^2$ in
%closed form.
%This yields an exact Gibbs step to sample directly from $p(\alpha,
%When we choose a $\operatorname{LG}(a,b)$ or $\operatorname{IG}(a,b)$, a Metropolis-Hasting step
%must be included...???

\textit{MCMC estimates}.
Once $H$ samples from the posterior distribution over $\thetavect$ are
available, it is possible to obtain Monte Carlo estimates of per-cell
risks by referring to (\ref{tau*:cella}):
\begin{eqnarray*}
\hat\tau_{1,k}^*&=&\frac{1}{H} \sum_{h=1}^H
\operatorname{Pr} \bigl\{ F_k = 1 | f_k = 1, \thetavect^{(h)}
\bigr\};\\
 \hat\tau_{2,k}^*&=&\frac{1}{H} \sum
_{h=1}^H E \biggl( \frac{1}{F_k}\Big |
f_k = 1, \thetavect ^{(h)} \biggr),
\end{eqnarray*}
which in turn lead to global risk estimates % $\tau_i^*$, $i=1,2$,
$
\hat\tau_i^*=\sum_{k=1}^K \hat\tau_{i,k}^*$, $i=1,2$. %, of

Fully Bayesian estimates of $\tau_{i},    i=1,2$, instead require
taking into account a further source of variability induced by the
randomness of the unobserved $F_1, \ldots, F_K$. %due to the fact that
%$F_1, \ldots, F_K$ are unknown.
In particular, observing that the terms $\tau_{i,k}$ in $\tau_{i}$, are
$\tau_{i,k}=\tau_{i,k}(f_k,\lambda_k,F_k)\    (i=1,2)$ %are
%functions
%of $f_k$, $\lambda_k$ and $F_k $
where %$\tau_i=\tau_i(F_1,\ldots,F_K, \mathbf{f})$, $i=1,2$,
$F_1,\ldots,F_K$ are unknown random quantities, with $F_k|\lambda_k
\sim\mathrm{Poisson}(\lambda_k)$, $k=1,\ldots,K$,
%fully Bayesian estimates of $\tau_{i}$ are obtained by adding a
%further step.
%us to. To perform posterior inference,
we consider values of $\lambda_k$'s drawn from their joint posterior
distribution and then values of $F_1,\ldots,F_K$ drawn from the
corresponding Poisson distributions.
%The relation $\tau_{ik}=\tau_{ik}(f_k,\lambda_k,F_k)   (i=1,2)$
This allows us to derive a sample of $\tau_{i,k}$, $i=1,2$, from which
it is possible to characterize the posterior distribution of global and
cell-specific risk values by standard Monte Carlo techniques.
Accounting for randomness of both groups of unobserved parameters
($\lambda_k$'s and $F_k$'s) has two important implications.
%%consequences. %distinguishing our results from the ones in the
%related literature.% we now discuss in the context of the previous
%literature.
%This approach takes into account the randomness of both groups of
%unobservable parameters ($\lambda_k$'s and $F_k$'s), with twofold
%consequences.
First, since a posteriori the $\lambda_k$'s are dependent on each
other, we avoid the unrealistic assumption %adopted in
underlying the second stage of the estimation procedure of \citet
{skinner:holmes}, where the cell risks are treated as if they were
independent. Second, since the uncertainty on the $F_k$'s is also
explicitly considered, we obtain risk estimates whose variability
depends on the variability of the $F_k$'s as well as the variability of
the $\lambda_k$'s and the association between $\lambda_k$'s. This
means, for instance, that
%, in the more general case,
our posterior variance of $\tau_1$,
$\operatorname{Var}(\tau_1|f_1,\ldots,f_K)=\operatorname{Var} (\sum_k ^K I(f_k=1)
I(F_k=1|f_k=1)|f_1,\ldots,f_K )$,
cannot be expressed in the form [\citet{rinott:shlomovar}]
%
%e13 #&#
\begin{equation}
\sum_k ^K I(f_k=1) \operatorname{Pr}
\{F_k=1|f_k=1\} \bigl(1-\operatorname{Pr}\{F_k=1|f_k=1
\} \bigr)
\end{equation}
%
%as in \citet{rinott:shlomovar}
because of the covariances of the $\lambda_k$'s. Moreover,
%we estimate these conditional
our variances, and the corresponding standard deviations (s.d.),
provided in
%Section~\ref{appl},
Table~\ref{tab:results3}, are %estimated by using
derived from
the posterior distributions of $\tau_i$, $i=1,2$, rather than by
plug-in. %a plug-in method.
%In order to keep the notation uncluttered, let $\theta$ denote the set
%of all model parameters,
%except for $F_1,..,F_K.$

%
%t1 #&#
\begin{table}
\caption{Estimated values of $\tau_1$ and $\tau_2$ by means of $\hat{\tau}_1$
and $\hat{\tau}_2$ (top panel) and $\hat{\tau}^*_1$ and $\hat{\tau
}^*_2$ (bottom panel) for the California and WHIP tables. Posterior
standard deviations in parentheses}\label{tab:results3} %True values of the global risks
%are $\tau_1 = 211$ and $\tau_2 = 499.820$ in the California table, and
%$\tau_1 = 915$ and $\tau_2 = 1948.056$ in the WHIP table. %with $K =
%2,160$,
%$\tau_1 = ??$ and $\tau_2 = ??$ in the California table, and $\tau_1 =
%39$ and $\tau_2 = 94.4$ in the small WHIP table.%with $K = 3,960$
\begin{tabular*}{\textwidth}{@{\extracolsep{\fill}}lcccc@{}}
\hline
& \multicolumn{2}{c}{\textbf{California}} & \multicolumn{2}{c@{}}{\textbf{WHIP}}\\[-6pt]
& \multicolumn{2}{c}{\hrulefill} & \multicolumn{2}{c@{}}{\hrulefill}\\
\textbf{Model}
& $\bolds{{\tau}_1=211}$ & $\bolds{{\tau}_2=499.8}$ & $\bolds{{\tau}_1=915}$ &
 $\bolds{{\tau
}_2=1948.1}$\\
\hline
 & $\hat{\tau}_1$ & $\hat{\tau}_2$ & $\hat{\tau}_1$ & $\hat
{\tau
}_2$ \\
(P$+$O) & {{0.0}} {{(0.1)}} & {{170.5}} {
{(1.4)}} & {{1180.7}} {{(33.2)}} & {{3322.2}}
{{(24.8)}} \\ % 2a
(P$+$I) & {{255.4}} {{(10.4)}} & {{518.8}}
{
{(7.6)}} & {{1184.9}} {{(23.7)}} & {{2289.9}}
{{(17.1)}} \\% 2c
(P$+$II) & {{253.9}} {{(11.1)}} & {{537.5}}
{
{(8.6)}} & {{958.4}} {{(22.4)}} & {{1996.2}}
{{(17.5)}} \\ % 2c
(NP$+$O) & {{700.0}} {{(232.1)}} & {{910.0}}
{
{(198.8)}} & {{2397.8}} {{(459.5)}} & {{3042.6}}
{{(405.1)}} \\% 3b
(NP$+$I) & {{217.0}} {{(12.2)}} & {{503.7}}
{
{(10.9)}} & {{1010.4}} {{(29.8)}} & {{2083.4}}
{{(28.3)}} \\
(NP$+$I) Emp & {{241.8}} {{(12.3)}} & {{528.8}}
{{(10.8)}} & {{970.2}} {{(32.7)}} & {
{2046.0}} {{(32.4)}} \\ [6pt]
& $\hat{\tau}^*_1$ & $\hat{\tau}^*_2$ & $\hat{\tau}^*_1$ & $\hat
{\tau
}^*_2$ \\
(P$+$O) & {{0.0}} {{(0.0)}} & {{170.6}} {
{(0.7)}} & {{1180.6}} {{(10.8)}} & {{3322.1}}
{{(10.8)}} \\ % 2a
(P$+$I) & {{255.3}} {{(3.3)}} & {{518.8}}
{
{(4.0)}} & {{1184.9}} {{(9.4)}} & {{2290.0}}
{{(9.7)}} \\% 2c
(P$+$II) & {{254.0}} {{(4.7)}} & {{537.6}}
{
{(5.5)}} & {{958.5}} {{(10.7)}} & {{1996.3}}
{{(11.9)}} \\ % 2c
(NP$+$O) & {{700.1}} {{(231.8)}} & {{910.1}}
{
{(198.7)}} & {{2397.8}} {{(458.8)}} & {
{3042.6}} {{(404.9)}} \\
(NP$+$I) & {{217.0}} {{(7.5)}} & {{503.7}}
{
{(8.7)}} & {{1010.3}} {{(21.9)}} & {{2083.4}}
{{(24.9)}} \\ % 3f
%NP UM & No & I & {{22.1}} {{(2.1)}} & {{52.0}} {
%{(2.4))}} & {{32.1}} {{(2.5)}} & {{86.5}} {
%{(3.2)}} \\ % 3e
(NP$+$I) Emp & {{241.7}} {{(7.4)}} & {{528.7}}
{{(8.5)}} & {{970.2}} {{(26.0)}} & {
{2046.0}} {{(29.6)}} \\[3pt]
II & {{250.4}} {{(--)}} & {{536.7}} {
{(--)}} & {{946.8}} {{(--)}} & {{1992.4}}
{
{(--)}} \\ % 2c
\hline
\end{tabular*}\vspace*{-5pt}
\end{table}

As mentioned in Section~\ref{intro}, the issue of nonexistence of MLE
(due to data being not fully informative about model parameters) is
also important in Bayesian analysis of log-linear models.
The vague prior we specify in Section~\ref{appl} for the fixed effects
replaces the information content lacking in the data with the
information contained in the prior. This prior is especially useful to
estimate the all two-way interactions model %and the all two-way
%interactions model (i.e. the same model without random effects)
%with and without parametric random effects
that we consider for comparison, as it makes the posterior information
matrix of $\betavect$ not rank deficient.
%. For each of these models, the posterior information matrix of $
% MAURIZIO E' VERA L'ULTIMA FRASE? DESCRIVE QUELLO CHE HAI FATTO?
This is the way we can avoid ad hoc additions of small positive
quantities to cells containing sampling zeroes. % It also is worth
\section{Applications}
% to Italian National Social Security Administration data and discussion
\label{appl}
To evaluate the performance of the proposed approach in practical
settings, we apply our nonparametric risk estimators to two tables with
different sizes and degrees of sparsity.
% and dimensions.
We consider data from the 5\% Public Use Microdata Sample of the U.S.
2000 Census for the state of California [IPUMS, Ruggles et al. (\citeyear{ru10})],
treating the set of individuals aged 21 and older as the population. We
also use data from the 7\% microdata sample of the Italian National
Social Security Administration (WHIP-Work Histories Italian Panel,  Laboratorio Revelli, Centre
for Employment Studies, \url{http://www.laboratoriorevelli.it/whip}), treated here as the population. In both cases we draw
random samples with fraction {$\pi= 0.05$}. The key variables
considered for the WHIP data are sex~(2), age~(12), area of
origin~(11), region of work~(20), economic sector~(4), wages guarantee
fund~(2), working position~(4) and firm size~(5), leading to a table of
$844\mbox{,}800$ cells, of which 5017 (0.59\%) are nonempty.
The California table comprises the following key variables: number of
children~(10), age~(10), sex~(2), marital status~(6), race~(5),
employment status~(3) and education~(5), for a total of $90\mbox{,}000$ cells,
of which 4707 (5.2\%) are nonempty. These variables are a subset of
those specified in \citet{mvreiter:jasa12} that we follow for
categorization of the key variables and selection of the reference
population; the latter excludes the presence of impossible, or
otherwise predetermined, combinations, that is, structural zeroes.
The expected cell probabilities ($\lambda_k$) in cells containing
structural zeroes are assigned a degenerate prior;\vadjust{\goodbreak} loosely speaking, this
% Given the previous definition of structural zero, such a degenerate
%prior
has to be interpreted as a ``conventional'' way to state that all such
cells have to be ignored in the fitting of the model
so that
%. Consequently,
they cannot bias estimates in the remaining ``nonstructural zero''
cells.%, whose parameters $\lambda_k$s are modeled by a log-linear

In the applications we focus on one of the nonparametric models
presented in Section~\ref{sec:2}, namely, the extension of the model proposed
by \citet{elamir:skinner}.
We examine several choices of the log-linear component describing the
fixed effects; in particular, we investigate a model with no fixed
effects, referred to as the overall mean model (O), the main effects or
independence %\footnote{More precisely, when there are structural
%zeroes the main effects model corresponds to a model of
%quasi-independence.}
model (I) and the all two-way interactions model (II).
For comparison we fit both the parametric (P) and nonparametric (NP)
random effects versions of the above-mentioned models.
For simplicity, hereafter, the above models will be identified by
labels denoting the selected modeling options, so, for instance, (NP$+$I)
is the nonparametric model with main effects only, and (P$+$II) and (II)
are the all two-ways interactions models with and without parametric
random effects, respectively.

Under the parametric specification P, the random effects $\phi$ are
modeled by a $\operatorname{LG}(a,b)$ prior, whereas
under the nonparametric specification NP, the random effects are
assumed to follow a distribution drawn from a DP whose base measure is
$\operatorname{LG}(a,b)$.
In both cases, the hyperparameters $(a, b)$ are fixed so that this is a
vague prior: $a=1$, $b=0.1$ ($b$ is the rate parameter).
Since we drop from $\betavect$ the overall effect $\beta_0$ to overcome
identifiability issues, $\beta_0$ is incorporated into the mean of the
random effects. Therefore, the assumption of \citet{elamir:skinner},
who take the mean of the Gamma distribution of the multiplicative
random effects $\omega$ to be 1, %the multiplicative random effects, %a
%Gamma with mean 1,
is compatible with ours: by fixing $a\neq b$, that is, a prior mean
that differs from 1, we simply allow for an overall effect.
For the components of $\betavect$ we assume independent and %a
reasonably vague Gaussian priors $N(0, 10)$. Finally, we take a
$\operatorname{Gamma}(1,0.1)$ prior on $m$.
All models are estimated by the fully Bayesian method\footnote{Suitably modified when estimating the
models (P$+$I) and (P$+$II).} described in
{Section~\ref{inf}}, with the exception of one nonparametric
independence model where the prior on the fixed effects is taken to be
degenerate at the MLE of $\xivect$, $\hat{\xivect}_{\mathrm{ML}}$.
We label the corresponding approach by (NP$+$I)~Emp to indicate that we
rely on empirical Bayes estimation in the presence of DP random
effects. %This approach is akin to the estimation strategy of
%generalization of the model in \citet{elamir:skinner}.
Note that the California table is free of structural zeroes, so that
the log-linear model with main effects only is in fact an independence,
that is, decomposable, model, and $\hat{\xivect}_{\mathrm{ML}}$ exists
since all observed unidimensional margins are positive. This is not the
case for the large WHIP table where the main effects model represents a
quasi-independence model.
%hypothesis.
Here we simply use the $\hat{\xivect}$ obtained by IPF (for which the
R routine converged within 15 iterations with a tolerance of
$10^{-8}$), assuming
%that
%this
it is the extended MLE.

In the implementation of the MCMC sampling, convergence of the chains
was checked using Gelman and Rubin's potential scale reduction factor
[$\hat{R}$; \citet{Gelman92}]\vadjust{\goodbreak} by running 10 parallel chains and
assessing that chains had converged when $\hat{R} < 1.1$ for all the
parameters.
According to this criterion, all chains converged within five thousands
iterations that were then discarded before running the chains for a
further $10\mbox{,}000$ iterations that were used to evaluate the risk scores.
%% (MINOR ISSUE REFEREE 2)

We note here that, for the (P$+$II) model, the $K \times q$ design matrix
associated with the log-linear model component is very large ($q >
10^3$ and $K \sim10^6$), which caused some difficulties when running
the adopted sampling scheme.
Indeed, each update of $\betavect$ requires evaluating and factorizing
a $q \times q$ matrix, leading to running times that are beyond
usability (weeks).
This is the main reason why we considered a subset of the variables in
the California table analyzed in \citet{mvreiter:jasa12}. For the
parametric models that are introduced for comparison we therefore
tested an alternative where we approximated the posterior distribution
over $\betavect$ by a Gaussian.
In particular, we carried out a Laplace Approximation, where the
approximating Gaussian has a mean equal to the mode of the posterior
distribution and the inverse covariance is equal to the negative
Hessian of the logarithm of the posterior density at the mode [\citet
{Tierney86}].
Computationally, this procedure has the following advantages.
First, the mode-finding procedure can be implemented in a way that it
does not require factorization or storage of large matrices, for
example, by feeding log-posterior and its gradient to standard
optimization routines.
%
%FRASE DA RIVEDERE: dobbiamo dire che l'approssimazione non d{\chr"E0}
%problemi; la frase sulla somiglianza con i loglin senza effetti
%casuali la sposterei.
%
Second, once the mode is located, drawing samples from the approximate
posterior over $\betavect$ requires that the $q \times q$ covariance
matrix is computed and factorized only once.
Interestingly, in cases where we could run the sampling from the
posterior over $\betavect$, we noticed that the risks obtained by the
approximate method were strikingly close to one another. % not
%substantially different from the corresponding log-linear model
%estimates without random effect obtained by running the IPF algorithm.
For this reason, the results that we report for the (P$+$I) and the
(P$+$II) models refer to the approximate method.

\begin{figure}

\includegraphics{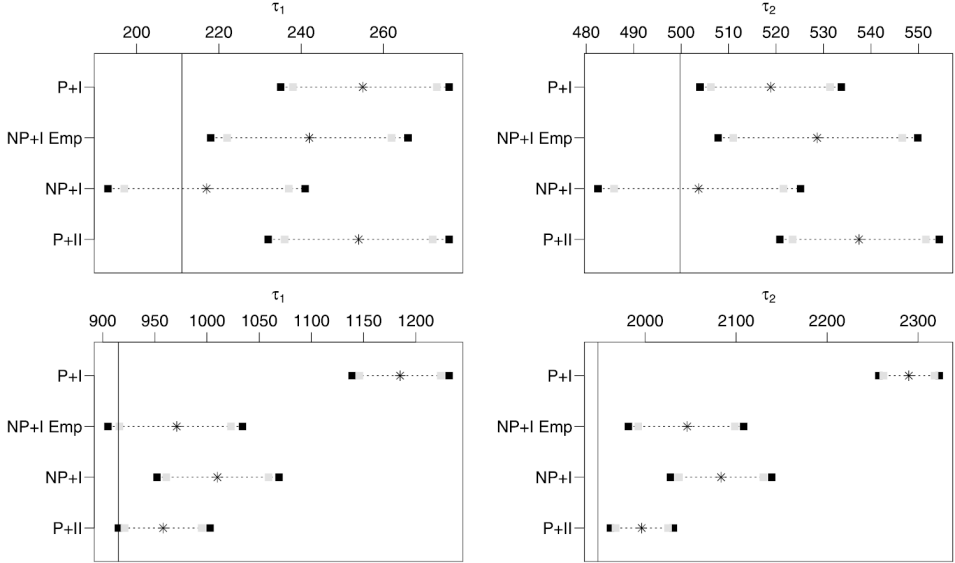}

\caption{Quantiles of the posterior distributions of ${\tau}_{1}$
(first column) and ${\tau}_{2}$ (second column)
under a subset of parametric and nonparametric models considered. First
(second) row refers to %the small Whip table, the second to the
%California 5\% sample, the third to the Whip 5\% sample. QUI ANDREBEBE
%SPIEGATO IN MODO DECENTE!
the California (WHIP) table. Gray squares: 5th, 95th percentiles; black
squares: 2.5th, 97.5th percentiles; stars:
median of the posterior distributions. Vertical segments represent the
true risks.}
\label{fig:quantili}
\end{figure}

%We also note that, because of the vague priors we adopted for $
%the fully Bayesian estimates of $\tau_1$ obtained under the models
%(P$+$I) and (P$+$II) can be considered roughly equivalent to the Empirical
%Bayes estimates obtained by \citet{elamir:skinner} when they follow
%the approach of \citet{skinner:holmes} for positive preliminary
%estimates of $\sigma^2$.
%
Table~\ref{tab:results3} reports true and estimated values of $\tau_1$
and $\tau_2$ (s.d. in parentheses) for six models formed by combining
different modeling options as described above. %Hereafter, the above
%models will be identified by labels denoting the selected modeling
%options. %aboveresulting from combination of the options just
%described.
In addition, risks obtained under the default log-linear model (II)
without random effects and fitted by the IPF are
%also
included for reference.
%Under the latter method point estimates do not differ substantially
%from the corresponding parametric random effects models, which also
%confirms the findings in \citet{elamir:skinner}.
First of all, the very small difference in the results under the (II)
and (P$+$II) models confirms the findings in \citet{elamir:skinner}.
Moreover, similar to what \citeauthor{mvreiter:jasa12}
[(\citeyear{mvreiter:jasa12}), page 1389] have
observed under their GoM models, %in Table~\ref{tab:results3}
point estimates $\hat{\tau}_1^*$ and $\hat{\tau}_2^*$ are nearly
identical to $\hat{\tau}_1$ and $\hat{\tau}_2$
% the ones in Table~\ref{tab:results3}, thereby confirming findings
%(1)-(3),
with smaller posterior standard deviations, since the former do not
take into account the variability of $F_k$'s. %(the slight variations
%are underlined).
%The standard errors are derived from the the posterior distribution
%of ${\tau}_i$ and $\tau_i^*$, $i=1,2.$
The 2.5th, 5th, 50th, 95th and 97.5th percentiles of the posterior
distribution of $\tau_i$, $i=1,2$, under a subset of the models
reported in Table~\ref{tab:results3}, are presented in Figure~\ref{fig:quantili} where models appear in order of complexity of the
log-linear specification and the solid vertical lines represent the
true risk values.

%f1 #&#

Inspection of Table~\ref{tab:results3}, and related Figure~\ref{fig:quantili},
confirms that the parametric all two-way interactions model (P$+$II)
outperforms the (P$+$I) model in the large table, %a satisfactory
%performance of the all two-way interactions model among parametric
%models in the large table,
which is in line with what was reported in the literature. If, however,
we include nonparametric models in the analysis, new and interesting
findings are as follows:
%% \begin{enumerate}
\begin{longlist}[1.]
\item[1.]  The potential of the DP prior for capturing association not
modeled by the fixed effects can be noticed
by comparing the results under the two models that, conditionally on
the random effects, rely on the exchangeability assumption, namely, the
parametric no fixed effects log-linear model (P$+$O) and its
nonparametric counterpart (NP$+$O).
% the parametric log-linear model that only contains the intercept,
%(P$+$O), and to
%%by observing the results under
%the nonparametric model without fixed effects, (NP+noF).
The latter is the model used in \citet{dorazio}. %for spatial
%heterogeneity in animal abundance.
%Both the above models rely on the unrealistic assumption of
%exachangeability of cells.
%the exchangeability assumption
%holds, and the pair $(\pi, K)$ is such that, in practice, no risks can
%be detected by $\tau_1$ under the (P$+$O) model in the California table.

\item[2.]  When risks are estimated by nonparametric models, the tendency
of risk estimates to decrease as the complexity of the model increases, shown
%associations underfitting-overestimation of the risks, and
%overfitting-underestimation, described in their article by
in \citeauthor{skinner:shlomo} [(\citeyear{skinner:shlomo}),
Table~1, going, in particular, from I to
II], can be observed in both California and WHIP tables
at a lower level, that is,
going from the (NP$+$O) model to the (NP$+$I) and (NP$+$I)~Emp models.

\item[3.]  The performance of the nonparametric independence model,
(NP$+$I)\break Emp, is roughly comparable to that of the parametric all two-way
interactions model, (P$+$II).
This means that the DP prior is able to capture the essential features
of heterogeneity without the need for additional terms (interactions)
in the vector of fixed effects. Considering, moreover, %Interestingly,
the good performance of the (NP$+$I) model %is also good
in the California table, we are induced to conclude that, in the
presence of DP random effects, the number of fixed effects required to
obtain reasonable global risk estimates is lower than in the parametric
case and less sensitive to the size of the table $K$. This is in line
with finding 2.
%increasing the dimensionality of the problem.
%nonparametric independence models, (P \& I) and (NP \& I), indicates
%that the a large improvement in risk estimates can be achieved at the
%cost of one additional unknown parameters ($m$), while (P \& II)
%requires thousands of additional parameters (all two-way interactions)
%with respect to (P \& I).

%nonparametric independence models, (P + I), (NP + I) and (NP $+$I )Emp,
%indicates the ability how the large improvement in risk estimates
%achieved by including DP random effects at the cost of one additional
%unknown parameter, $m$.

%(Non so se aggiungere qui che, dunque, NP$+$I e' un naturale modello di
%partenza per la ricerca di un modello ottimo con procedure forward o
%backward, il che significa che il problema della scelta del modello,
%seppur non affrontato qui, e' sensibilmente ridimensionato all'origine
%nella sua difficolta'. - Nella risposta ai referee potremmo dire che
%sar{\chr"E0} oggetto di lavoro futuro).
% the fully Bayesian estimates of $\tau_1$ obtained under the models
%(P$+$I) and (P$+$II) can be considered roughly equivalent to the Empirical
%Bayes estimates obtained by \citet{skinner:holmes} when their
%preliminary estimate of $\sigma^2$ is positive.
\item[4.]  Although we do not specifically address the challenging
problem of model choice, our approach may contribute to lessen its
scale and complexity.
Indeed, the (NP$+$I)~Emp model can be taken as the initial model in a
forward model selection procedure. The significant reduction of the
space of adjacent models that need to be examined at each step would mitigate
%probably, we indirectly contribute to reduce the degree of difficulty
%of this problem. For instance, taking our (NP$+$I) default model,
%instead of the standard all two-ways interactions model without random
%effects, as the initial model in a forward model selection procedure
%may mitigate
the difficulties associated with model choice. This point will be
explored in future work.
\end{longlist}

By comparing parametric and nonparametric independence models, (P$+$I),
(NP$+$I) and (NP$+$I)~Emp, Figure~\ref{fig:quantili} allows us to see how
strongly DP random effects integrate into a log-linear model with main
effects only and contribute to improving global risk estimates even for
the large WHIP table for which the fit of the parametric independence
model is particularly poor.

\begin{figure}

\includegraphics{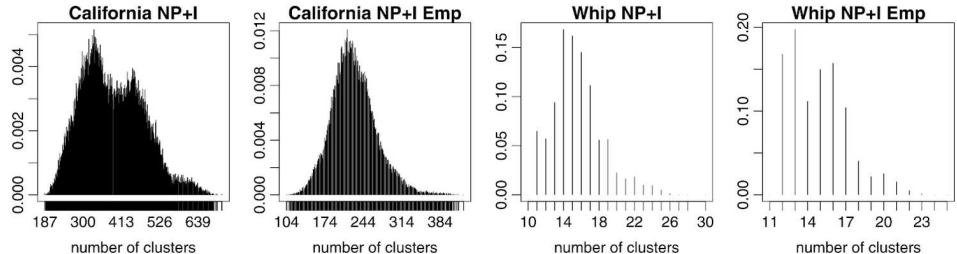}

\caption{Posterior distribution of the number of clusters for the
California and WHIP tables when using the (NP$+$I) and (NP$+$I) Emp models.}
\label{fig:numclassi}
\end{figure}

To appreciate the role played by the clustering mechanism induced by
the DP, Figure~\ref{fig:numclassi} provides a representation of the
posterior distribution of the number of clusters under the proposed
(NP$+$I) and (NP$+$I)~Emp models.
%% Suggerimento Maurizio
There is a striking difference between the distribution of the number
of clusters for the California and WHIP tables.
The fact that in the California table the number of clusters is large
seems to reflect the ability of the (NP$+$I) model to perform extremely
well in the estimation of risk.
In the case of the WHIP table, the introduction of the DP distributed
random effects, although significantly improving on the estimation of
risk with respect to the (P$+$I) model,
does not completely account for the lack of fit.

%seems to offer risk estimates that are slightly off the true values.
%The interesting feature of this work, is that the clustering patterns
%that emerge from the analysis are completely data-driven.

%We can notice that for the %smaller
% California table the number of clusters is substantially larger and
%therefore a much larger number of patterns of dependence is taken into
%account.
%When the the fixed effects component of the log-linear model
% %component
% is not estimated simoultaneously, the method does not exploit the
%interaction between its parametric fixed effects %parametric
% and nonparametric random effects components and the number of
%clusters is lower.
%For the large table the worse performance of the (NP$+$I) model could be
%ascribed to a less powerful exploitation of the nonparametric
%clustering feature of the model.QUESTA FRASE QUI NON LA SCIVEREI
%PERCHE' NON E' COERENTE CON QUELLA DI SOPRA E CON IL COMPORTAMENTO DI
%(NP$+$I)emp, E FORSE ANCHE QUELLA DI SOPRA E' RISCHIOSA DA SCRIVERE:
%SONO I DATI CHE HANNO DECISO COSI'.
%

%f2 #&#

%f3 #&#
\begin{figure}

\includegraphics{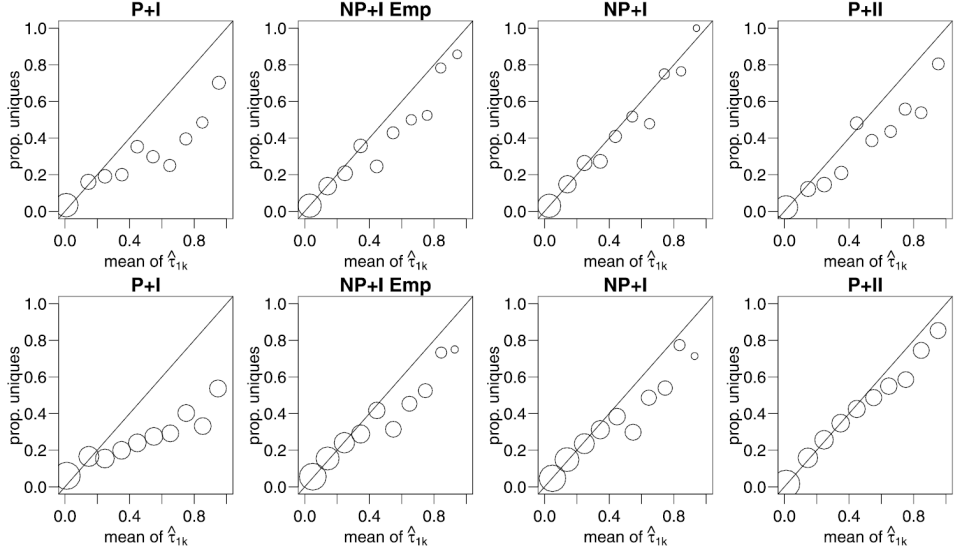}

\caption{Proportion of population uniques plotted against the average
estimated risk $\hat{\tau}^*_{1,k}$, for cells categorized into 10
equal-width intervals according to the values of $\hat{\tau}^*_{1,k}$.
The size of the plotting points depends on the number of cells in each
interval. First line: California table; second line: WHIP table.}
\label{fig7:tau1}
\end{figure}

% \begin{figure}[t]
% \centering
% \includegraphics[width=.24
% \includegraphics[width=.24
% \includegraphics[width=.24
% % %&
% %% \caption{Posterior distribution of the number of clusters for the
%California (top row) and the WHIP table (bottom row) when using the
%NP$+$I model (left) and the NP$+$I Emp (right), respectively.}
%
% \end{figure}

For the California table we also explored the frequentist
properties of our approach through a simulation study comprising 100
samples, where we evaluated the frequentist coverage of the credible
intervals based on the 2.5th and 97.5th percentiles of the posterior
distribution of $\tau_i, i = 1,2$. We observed that, under the (NP$+$I)
model, all of them include the true value of $\tau_1$ and 76 include
the true value of $\tau_2$.
%
%By exploiting such a simplification,

\begin{figure}

\includegraphics{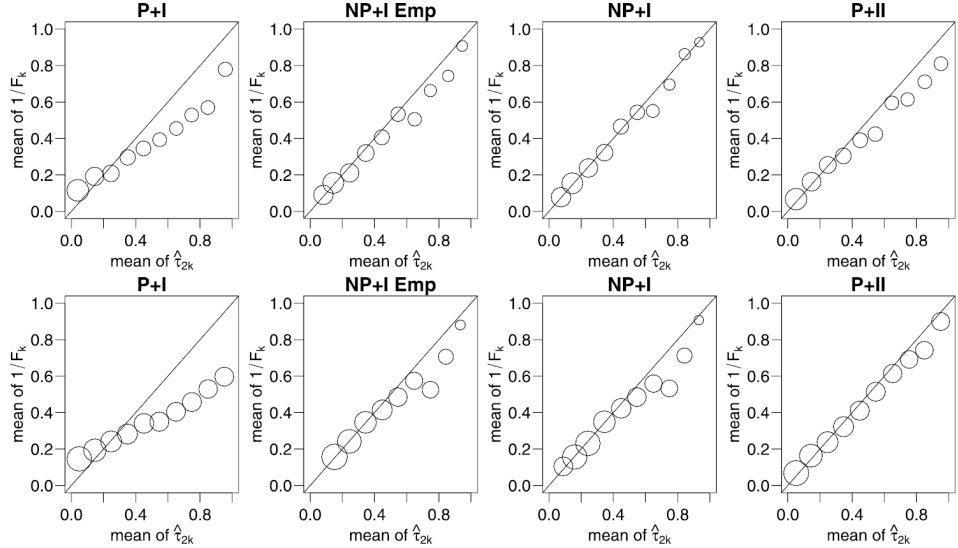}

\caption{Mean of $1/F_k$ against the mean of the estimated risk $\hat
{\tau}^*_{2,k}$, for cells categorized into
10 equal-width intervals according to the values of $\hat{\tau
}^*_{2,k}$. The size of the plotting points depends on
the number of cells in each interval. First line: California table;
second line: WHIP table.}
\label{fig8:tau2}
\end{figure}

In the rest of this section we explore the behavior of per-cell risk
estimates by using, for simplicity, $\hat{\tau}_{1,k}^*$ and $\hat
{\tau
}_{2,k}^*$. In Figure~\ref{fig7:tau1}, for a subset of the models
presented in Table~\ref{tab:results3}, and proceeding as in Figure~4 of
\citet{forster:webb}, we plot the proportion of population uniques
against the average value of $\hat{\tau}^*_{1,k}$, for cells
categorized into 10 equal-width intervals according to the values of
$\hat{\tau}^*_{1,k}$. Visual assessment of the relative proximity to
the diagonal gives an idea of how accurately each model can predict
population unique cells. %The size of the plotting point is
%proportional to the number of cells in each of the intervals.
Similarly, in Figure~\ref{fig8:tau2}, as in \citet{elamir:skinner}, we
plot the mean of $1/F_k$ against the mean of the estimated risk $\hat
{\tau}^*_{2,k}$ after grouping cells into 10 intervals according to the
values of $\hat{\tau}^*_{2,k}$. %CoComparing the (NP$+$I) and (P$+$II)
%models, in all these plots we can see that the (NP$+$I) model tends to
%slightly overestimate the mean or mediane risks in cells with large
%population frequencies, while in cells with small population
%frequencies both models tend to underestimate the corresponding risks
%(CONTROLLARE).

% \centering
% %\includegraphics[width=0.99\textwidth]{tsplots_tau2k_tab1k.pdf}
% % era {tsplots_tau1k_PII_vs_I_insieme}
%%\includegraphics[width=0.24
%
%
%
% \\
%
%
%
% \\
%
%
% \end{figure}

%f4 #&#

\begin{figure}[b]

\includegraphics{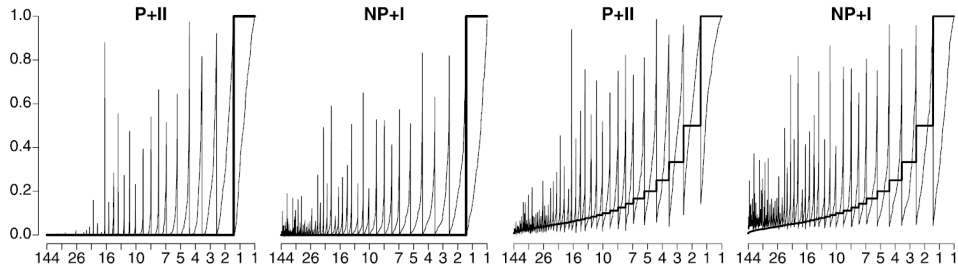}

\caption{Comparison of risk estimates $\hat{\tau}^*_{1,k}$ (first two
plots) and $\hat{\tau}^*_{2,k}$ (last two plots)
for sample unique cells under the (P$+$II) and (NP$+$I) models for the
California table. Bold lines represent the true risks.}
\label{fig:tau:PII vs NP}
\end{figure}

In Figure~\ref{fig:tau:PII vs NP} we compare per-cell risk estimates
$\hat{\tau}^*_{i,k}$ and true risks (bold lines) for $i=1, 2$,
respectively. We consider estimates from the California table for which
the (NP$+$I) model outperforms the parametric model (P$+$II) and the
parametric independence model (P$+$I). Cells containing sample uniques
are arranged in increasing order of the true per-cell risk; in turn,
for each level of the true per-cell risk, estimates are arranged in
decreasing order of population cell size and increasing order of
estimated risk. This allows us to observe overestimates and
underestimates in all cells under the two models under examination. By
drawing cutoff points (not included) in the first two plots
%top row
of
the figure, we can also visualize the corresponding false positive
%%cells
and false negative cells. %We can conclude that the good perfomance of
%the (NP$+$I) model in global risk estimation can be explained mainly in
%terms of improved estimates in cells with intermediate population
%frequencies, while in cells with very large frequencies or with a
%frequency of 1 (population uniques) the (P$+$II) model tends to produce
%better results.
We can conclude that the (NP$+$I) model improves risk estimates $\hat
\tau
_{1,k}$ in cells with intermediate population frequencies, while in
cells with extreme (very large or 1) population frequencies, the (P$+$II)
model tends to produce better results at the cell level; however, this
is not sufficient for the (P$+$II) model to outperform the (NP$+$I) model
in the estimation of the global risk.
This fact is even more apparent when inspecting the last two plots in
%bottom panel of
Figure~\ref{fig:tau:PII vs NP}. The results just analyzed indicate
that, compared to the all two-way parametric random effects log-linear
model, the proposed approach does not produce uniformly better per-cell
risk estimates. While in this paper we have mainly focused on measures
of global risk, the specific problem of per-cell risk estimation could
be tackled in a different way, that we plan to explore in future work.
%% We also note that from the results reported in the literature
%estimation of $\tau_2$ appears to be more challenging than estimation
%of $\tau_1$. %FIGURE...sui TAU2 di cella e suggerisce FRASE DA
%CALIBRARE BENE

%f5 #&#

\section{Computational aspects and comparison with other approaches}
\label{sec:comput}
In this section we discuss computational costs and applicability to
large tables of our proposal, in comparison with other approaches in
the recent literature related to our problem.

%As already discussed, in large tables maximum likelihood estimation of
%standard log-linear models and model search become highly challenging,
%as the parameter space a number of parameters may no longer be
%%%%%identifiable due to sparsity.
%%%%As already discussed, in large and sparse tables maximum likelihood
%%%%estimation of standard log-linear models-(II), in particular in our
%%%%application- and model search becomes highly challenging, as the
%%%%parameter space %may
%%%%quickly explodes and a number of parameters may result in being unidentifiable
%%%%%no longer be identifiable
%%%%due to sparsity. %This is the reason why, for the California data, we
%%%%%considered a subset of the table presented in \citet{mvreiter:jasa12}.
%%%%% (MI SEMBRAVA UN PO' DISONESTO PRIMA)
As already discussed, in large and sparse tables maximum likelihood estimation
of standard log-linear models---in particular the all two way interaction
model~(II) in our application---and model search become highly challenging, as the
parameter space quickly explodes and a number of parameters may result in
being unidentifiable due to sparsity.
Assigning a prior to these parameters and carrying out
Maximum-A-Posteriori (MAP) estimation instead of Maximum Likelihood,
allowed us to somewhat work around this problem.
However, locating the mode of the posterior distribution of the
parameters requires employing iterative search algorithms that are
computationally intensive and potentially slow to converge.
%Also, quantification of uncertainty of parameters becomes heavily
%affected by the choice of the prior on the parameters, which is not
%ideal to reliably quantify uncertainty in risk estimates.
%The computational performance and the possible computational issues of
%our nonparametric approach, which is an independence log-linear models
%with nonparametric random effects, depend on the interplay of two
%different elements,
%%aspects,
%namely: (i) estimation of the parametric fixed effects; (ii)
%estimation of the nonparametric random effects.
%
%As to (i), it is the number of fixed effects to be estimated that
%determines the computational scale of the problem, rather than the
%table size (although they are clearly related). This is because,
%conditional on the random effects, we focus on simple independence
%models. When the number of fixed effects is very large, storing of
%information matrices might be challenging; in that case we suggest use
%of the Empirical Bayesian version of the nonparametric independence
%model. %, where the prior on the fixed effects is taken to be
%degenerate at the MLE of $\xivect$, $\hat{\xivect}_{\mathrm{ML}}$.
%This approach %, akin to the estimation strategy of
%is an appealing alternative, since it relies on IPF that converges in
%at most two steps when the model is decomposable. \\
%
%I PRECEDENTI DUE PARAGRAFI POTREBBERO SEMPLIFICARSI IN:\\

Vice versa, the computational performance of our nonparametric
independence models depends on the interplay of two different elements,
namely, (i) estimation of the parametric fixed effects $\betavect$; and
(ii) estimation of the nonparametric random effects.
%As to (i), it is the number of main effects that determines the
%computational scale of the problem (cubic in the size of $\beta$).
%Although this is a great simplification of the problem, when the
%number of main effects is very large, storing of information matrices
%might be challenging; in that case we suggest use of the Empirical
%Bayesian version of the nonparametric independence model. %, where the
%prior on the fixed effects is taken to be degenerate at the MLE of $
%This approach %, akin to the estimation strategy of
%is an appealing alternative, since it relies on IPF. Note also that it
%is in line with the estimation strategy originally proposed by
%generalization. This strategy that is applied in Section~\ref{appl} to
%generalize the model in \citet{elamir:skinner}; although neglecting
%the variability of the fixed effects, it incorporates other sources of
%uncertainty such as the population frequencies.\\
% FINE VARIAZIONE. \\
% \end{comment}
As to (i), it is the number of main effects that determines the
computational scale of the problem, that, albeit cubic in the size of
$\betavect$, remains much smaller than the table size. Nonetheless,
when the size of $\betavect$ is very large, storing of information
matrices might be challenging; in that case we suggest use of the
Empirical Bayesian version of the nonparametric independence model.
This approach, akin to the estimation strategy of \citet
{skinner:holmes}, is an appealing alternative, since it relies on IPF
that converges in at most two steps in decomposable models.

(ii) is related to the allocation of random effects; the proposed
nonparametric methods scale linearly with the number of cells, which
makes our proposal suitable for applications to large tables.
%Unfortunately,
Although it is not possible to provide any guarantees on convergence
speed of the MCMC approach to the posterior distribution over the
parameters,
%Having said that,
in all tables that we studied in this work, we found that convergence
of the chains was reached after a few thousand iterations. % e
%----spazio per dire tutto quello che di buono succede con i $\lambda$.

By using a log-linear representation of the latent class model (LCM),
our (NP$+$I) model and the LCM recently applied by \citet
{manriquevallier:reiter:13} can be shown to rely on the same basic
assumptions, that is, independence of the key variables conditionally
on an unobserved variable $S$ and a prior for such unobserved variable $S$
somewhat related to the Dirichlet process [see also \citet{Si:Reiter}].
%%The latter assumption, however, is applied in different ways in the
%two models implying a different clustering and different sampling
%schemes.
%It is worth noting that, using a log-linear representation of the
%latent class model (LCM), our (NP$+$I) model and the LCM recently
%applied by \citet{manriquevallier:reiter:13} can be shown to rely on
%the same basic assumptions, i.e. independence of the key variables
%conditionally on an unobserved variable S, and a prior for such
%unobserved variable S somewhat related to the Dirichlet process
%[see also \citep{Si:Reiter}, ].
The latter assumption, however, is applied at the level of individuals
(through an individual latent class $Z$ whose prior is a finite stick
breaking process) in the LCM, while it is applied at the level of cells
(via the cell-specific DP random effect $\phi$) in the (NP$+$I) model.
This implies different allocations to clusters and different sampling
schemes in the two cases. Practical consequences are that as the sample
size increases, the (NP$+$I) model does not require any additional
computational costs, while it scales as discussed above with the number
of cells. Vice versa, the LCM scales easily with the number of cells,
as emphasized in \citet{manriquevallier:reiter:13}, but has to sustain
a nonnegligible computational cost as the sample size increases. This
may be an advantage of our method, as %the samples usually dealt by
%Statistical Agencies, if not larger, are of comparable sizes with
%respect to the ones we chose for our applications ($n=57,547$;
%$n=40,122$ for the California and WHIP data, respectively). %I
in the practice of Statistical Institutes the sample sizes are often
much larger than those considered in the literature based on LCMs. Note
that while in our applications the sampling fraction is higher than
what could commonly be used in practice, the absolute size ($n=57\mbox{,}547$;
$n=40\mbox{,}122$ for the California and WHIP data, resp.) is the same
order of magnitude of many surveys on individuals conducted, for
instance, by the Italian National Statistical Institute. A second
practical issue relates to structural zeroes, which, at the level of
cells, are very simply managed in our nonparametric approach (by a
degenerate prior on those cells), while they require a specific
technique in the LCM, that is, the one introduced by \citet
{manriquevallier:reiter:13}.
%in an extremely large table including structural zeroes.
This also means that our approach has the same advantages mentioned by
\citet{manriquevallier:reiter:13}, such as applicability to variables
with skip patterns or when certain combinations have been effectively
eliminated from the sample by design.

%Finally, \citet{Si:Reiter} consider a latent class model for
%incomplete data - instead of incomplete tables - and adopt a DP prior
%for the latent variable $Z$, thereby avoiding to fix a priori the
%number of classes \citet{manriquevallier:reiter:13}. Then, ignoring
%classes with negligible mass, they describe the inferences under their
%model as averages over models with different finite number of classes.
%Analogously, the results under our nonparametric models can be
%interpreted as averages over mixed effects log-linear models with
%different clusterizations of parametric random effects.

In conclusion, (NP$+$I) and LC models are built on the same basic
ingredients though implemented in different ways, thereby producing the
different advantages and disadvantages---in terms of scalability,
structural zeroes and applicability---just summarized.

The actual computational times associated with our proposal clearly
depend on the size of the table to be analyzed. % It is difficult to
%quantify the typical table size in real applications of disclosure
%risk estimation, for the reasons that we explain next. %, and the
%available IT tools.
Recent developments on Bayesian LCMs show that they can deal with
extremely large tables, in the order of $10^{40}$ as illustrated by
\citet{Si:Reiter} for a multiple imputation problem. Being able to
treat extremely large tables in short computational times is
undoubtedly important. Although ``Big Data'' issues are likely to have
an impact in the context of disclosure risk estimation (in terms of
disclosure scenario and type and number of key variables), % it is
%difficult to quantify the typical table size in real applications of
%disclosure risk estimation, and
we deem that tables of the above size may be less common than in other
related fields. Indeed, when the number of cells is much higher than
the population size, the average population cell size $N/K$, whatever
the sample, is very low. Under such circumstances Statistical
Institutes may judge releasing information on the key variables at that
level of detail too risky and may prefer to recode/merge the key
variables and/or decrease their detail before proceeding to assess the
risk formally through a suitable statistical model.
\section{Final comments}
% to Italian National Social Security Administration data and
%discussion
\label{disc}
%differently from
%differently from
In this article we investigated the role of random effects in
log-linear models for disclosure risk estimation. %As in
We show in theory and through real data applications that modeling the
random effects nonparametrically does improve upon the log-linear
model, because it allows to simplify to a large extent the
%In this article we re-introduce the random effects in log-linear
%models for disclosure risk estimation and show that a Bayesian
%nonparametric specification of their prior greatly simplifies the
structure of fixed effects required to achieve good risk estimates.
%
%Quoting \citet{fienberg2007}, ``the number of possible patterns of
%zero counts invalidating the MLE exibits an exploding behaviour as the
%number of classifying variables or categories grows''. In this respect,
Therefore, the utility of our nonparametric approach increases with
the size and the degree of sparsity of the table, %increases,
since problems with nonestimable parameters in fixed effects
log-linear models %including high order terms
%the more complex log-linear interactions model
increase disproportionately with the number of terms included. Quoting
\citet{fienberg2007}, ``the number of possible patterns of zero counts
invalidating the MLE exhibits an exploding behavior as the number of
classifying variables or categories grows.''

Unlike parametric random effects models, for each cell our
nonparametric models combine learning from two types of neighborhoods,\vadjust{\goodbreak}
one driven by the fixed effects, and the other driven by the data and
implied by the clustering of the random effects.%Indeed the number of
%possible patterns of zero counts invalidating the MLE exhibits an
%exploding behavior as the number of classifying variables or
%categories grows \citep{fienberg2007}. %reduces the number of fixed
%effects required to obtain reasonably good estimates of risks. %need
%of this assumption i.
%Indeed,
% The posterior distribution of the number of clusters in the (NP$+$I)
%model for the California table is provided in Figure~\ref{fig:numclassi}.
%QUESTO FORSE {\chr"E8} DA SPOSTARE NEL PAR 4.
% To appreciate the clustering effect, Figure~\ref{fig:numclassi}
%provides a representation of the posterior distribution of the number
%of clusters under the proposed (NP$+$I) model.
%Moreover we show that the above assumption simplifies to a large
%extent the
%In this article we re-introduce the random effects in log-linear
%models for disclosure risk estimation and show that a Bayesian
%nonparametric specification of their prior greatly simplifies the
%structure of fixed effects required to achieve good risk estimates.
%In this respect, the utility of our nonparametric approach increases
%with the degree of sparsity of the table, %increases,
%since problems with nonestimable parameters in fixed effects
%log-linear models including high order terms
%the more complex log-linear interactions model
% increase disproportionately with the number of terms added. Indeed
%the number of possible patterns of zero counts invalidating the MLE
%exhibits an exploding behavior as the number of classifying variables
%or categories grows \citep{fienberg2007}.

Interestingly, in the applications the empirical Bayesian version of
our (NP$+$I) model emerges as the nonparametric equivalent of the
parametric model (P$+$II), indicated in the literature as the default
approach in risk estimation. This evidence is found in tables with
rather different structures and dimensions. Moreover, in the analysis
of the California data set
%indicates that
the (NP$+$I) model %may
greatly improves the performance of the parametric model in terms of
global risk estimation.

The striking impact of the inclusion of DP random effects in the (P$+$I)
model indicates that enlarging the simple (NP$+$I)~Emp model by adding a
few interaction terms can be expected to produce satisfactory results.
Even if we do not address the issue of model selection, the previous
remark opens the door to a model search approach that takes our
(NP$+$I)~Emp model as the starting point, thus lessening the scale and
complexity of the problem, since the space of adjacent models to be
examined is significantly reduced. %A computationally more efficient
%model selection strategy would be to adopt the Empirical Bayesian
%(NP$+$I Emp) version of the former model.
%Limiting the search to decomposable models would facilitate the
%problem under nonexistent MLEs.

We emphasize that the previous ones are general results, that is, a
reduction in the number of fixed effects in the presence of DP random
effects---with the mentioned benefits in terms of estimability in
sparse tables and simplification of model search---can be expected in
different applications of log-linear models, not only in disclosure
risk estimation.
%% Our approach is cast in a Bayesian framework. We remark that in a
%disclosure limitation context, the sample to be released is unique and
%fixed, and conditioning on the observed data is more coherent than
%referring to repeated sampling schemes.
%% Moreover

Having adopted a fully Bayesian approach allowed us to account for all
sources of uncertainty (about $\lambda_k$'s, $F_k$'s) in the estimation
of risk.
%% The impact of such an approach can be noticed when considering
%direct generalizations of previous works, like the estimates of $
%(P$+$II). These are the fully Bayesian versions of the empirical Bayes
%estimates obtained by \citet[Sect.~3.3]{elamir:skinner}. %when they
%follow the approach of \citet{skinner:holmes} for positive preliminary
%estimates of $\sigma^2$.
%% Because of the vague priors adopted for the fixed effects in (P$+$I)
%and (P$+$II) models, we can expect numerical agreement between their
%estimates and the empirical Bayes ones obtained by \citet[Sect.~3.3]{elamir:skinner}, while providing a quantification of uncertainty.
%% in the estimates.
In (P$+$I) and (P$+$II) models, this is an advantage compared to the
empirical Bayes procedure by \citeauthor{elamir:skinner}
[(\citeyear{elamir:skinner}), Section~3.3], even
though we can expect numerical agreement between their estimates
because of the vague priors adopted for the fixed effects.
%% The impact of such an approach can be noticed when considering
%direct generalizations of previous works, like the estimates of $
%(P$+$II). These are the fully Bayesian versions of the empirical Bayes
%estimates obtained by \citet[Sect.~3.3]{elamir:skinner}.
%% Because of the vague priors adopted for the fixed effects,
%% we can expect numerical agreement between the approaches, with the
%difference that our fully Bayesian estimates account for all sources
%of uncertainty (about $\lambda_k$'s, $F_k$'s).
%% As to our (NP$+$I)~Emp model, that replicates in a nonparametric
%context the estimation strategy of Skinner and Holmes (1998), although
%neglecting the variability of the fixed effects, it incorporates other
%sources of uncertainty such as the population frequencies.
As to our (NP$+$I)~Emp model, which replicates in a nonparametric context
the estimation strategy of Skinner and Holmes (\citeyear{skinner:holmes}), although it
neglects the variability of the fixed effects, it incorporates other
sources of uncertainty, such as the population frequencies.
Although our approach generalizes existing models mentioned in
Section~\ref{sec:2}, there are important differences from the previous
literature, including \citet{rinott:shlomovar}, as our risk estimates
are endowed with unconditional (posterior) variances and we can also
produce credible intervals, that is, posterior probability intervals.
%%These are important differences from the previous literature,
%including \citet{rinott:shlomovar}, as our risk estimates are endowed
%with unconditional (posterior) variances and we can also produce
%credible intervals, i.e. posterior probability intervals. %This
%represent a novel contribution to the literature on disclosure risk
%based on log-linear models. %presented in Figure~\ref{fig:quantili}.
%%To address a Referee's question about the frequentist coverage
%properties of our intervals, we have explored 100 samples drawn from
%the
%We have inspected frequentist coverage properties of our nonparametric
%approach through a simulation study for the California data for which
%our models gave a satisfactory fit. The results obtained indicate a
%very good coverage of our (NP$+$I) estimator.

As regards the assumptions underlying our Bayesian models, all of them
are explicit and more flexible than the ones underlying a log-linear
model without random effects. Indeed, we have selected vague priors and
modeled the random effects nonparametrically, which is a further
relaxation of the hypotheses. % It is worth mentioning that the use of
%a prior for the fixed effects $\betavect$ is useful to overcome the
%already mentioned problems of existence of the MLE.
%% The computational burden of our method is balanced by the resulting
%substantial simplification of the fixed effects component of the model
%and by the advantages offered to overcome the difficulties encountered
%by standard estimation methods with complex models in sparse tables.
%% Having adopted a fully Bayesian view, we have introduced prior
%distributions for all the model parameters, selecting vague priors to
%avoid strong assumptions. The use of a nonparametric specification of
%the random effects is a further relaxation of the hypotheses upon
%which our model is based. All assumptions underlying our model are
%explicit and more flexible than the ones underlying a log-linear model
%without random effects. It is worth mentioning that the use of a prior
%for the fixed effects $\betavect$ is useful to overcome the already
%mentioned problems of existence of the MLE.
%% The computational burden of our method is balanced by the resulting
%substantial simplification of the fixed effects component of the model
%and by the advantages offered to overcome the difficulties encountered
%by standard estimation methods with complex models in sparse tables.

%we provide unconditional measures of varibility
% based on posterior summaries and credible intervals for sample
%disclosure risk measures.
While in this paper we have mainly focused on measures of global risk,
the applications indicate that, compared to the all two-way parametric
random effects log-linear model, the proposed approach does not produce
uniformly better per-cell risk estimates even when the global risk
estimates under the (NP$+$I) model outperform those obtained under the
(P$+$II) model.
The specific problem of per-cell risk estimation could be tackled in a
different way, that we plan to explore in future work.
\section*{Acknowledgments}
The authors wish to thank the Associate Editor and the anonymous reviewers for their valuable and constructive comments.\vadjust{\goodbreak}

% imsref loaded by akundreckaite, 2015-02-12 15:15:06

%

%suskaldyti doi

\printaddresses
\end{document}